\documentclass[reprint, NumberedRefs]{ARXIVJASA}

\usepackage{mathtools}

\renewcommand{\t}[2]{\tensor*{#1}{#2}}
\newcommand{\mbf}[1]{\mathbf{#1}}
\newcommand{\mbb}[1]{\mathbb{#1}}
\newcommand{\mrm}[1]{\mathrm{#1}}
\newcommand{\tbf}[1]{\textbf{#1}}
\newcommand{\tit}[1]{\textit{#1}}
\newcommand{\mcl}[1]{\mathcal{#1}}

\newcommand{\til}[1]{\widetilde{#1}}
\newcommand{\sot}{\mrm{SO}(3)}
\newcommand{\tthre}{\mbf{T}^3}
\newcommand{\bsquare}{$\blacksquare$}

\newcommand{\mi}[0]{\mrm{i}}   
\newcommand{\Acoef}[2]{        
    \t{A}{_{#1}^{#2}}
}
\newcommand{\acoef}[2]{        
    \t{a}{_{#1}^{#2}}
}

\newcommand{\ahcoef}[2]{        
    \t{\widehat{a}}{_{#1}^{#2}}
}

\newcommand{\bcoef}[2]{        
    \t{b}{_{#1}^{#2}}
}

\newcommand{\Ccoef}[2]{        
    \t{C}{_{#1}^{#2}}
}

\newcommand{\WD}[3]{\t{D}{_{#1}^{#2 #3}}}

\newcommand{\wwo}{(2\omega;\omega,\omega,0)}
\newcommand{\wow}{(2\omega;\omega,0, \omega)}
\newcommand{\oww}{(2\omega;0,\omega,\omega)}

\usepackage[utf8]{inputenc}
\usepackage{amssymb}
\usepackage{cleveref}
\usepackage[section]{placeins}
\usepackage{tensor}
\usepackage{amsthm}
\usepackage{tikz}
\usepackage{enumitem}

\theoremstyle{definition}
\newtheorem{theorem}{Theorem}[]

\theoremstyle{definition}
\newtheorem{definition}[theorem]{Definition}

\theoremstyle{definition}

\theoremstyle{definition}
\newtheorem{lemma}[theorem]{Lemma}

\theoremstyle{definition}
\newtheorem{corollary}[theorem]{Corollary}

\theoremstyle{definition}
\newtheorem{remark}[theorem]{Remark}

\begin{document}

\title[Compressive Spherical Field Measurements]{On Grid Compressive Sampling for Spherical Field Measurements in Acoustics$^{\star}$}


\author{Marc Andrew Valdez}
\email{mvaldez@mines.edu}
\altaffiliation{Also at: Dept.~of Electrical Engineering, Colorado School of Mines, Golden, CO 80401, USA\\ \\$^{\star}$This work was partially supported by U.S.~government, not protected by U.S.~copyright.}

\author{Alex J.~Yuffa}
\affiliation{National Institute of Standards and Technology, Boulder, CO 80305, USA.}

\author{Michael B.~Wakin}
\affiliation{Dept.~of Electrical Engineering, Colorado School of Mines, Golden, CO 80401, USA.}









\begin{abstract}
    We derive a compressive sampling method for acoustic field reconstruction using field measurements on a predefined spherical grid that has theoretically guaranteed relations between signal sparsity, measurement number, and reconstruction accuracy. This method can be used to reconstruct band-limited spherical harmonic or Wigner $D$-function series (spherical harmonic series are a special case) with sparse coefficients. Contrasting typical compressive sampling methods for Wigner $D$-function series that use arbitrary random measurements, the new method samples randomly on an equiangular grid, a practical and commonly used sampling pattern. Using its periodic extension, we transform the reconstruction of a Wigner $D$-function series into a multi-dimensional Fourier domain reconstruction problem. We establish that this transformation has a bounded effect on sparsity level and provide numerical studies of this effect. We also compare the reconstruction performance of the new approach to classical Nyquist sampling and existing compressive sampling methods. In our tests, the new compressive sampling approach performs comparably to other guaranteed compressive sampling approaches and needs a fraction of the measurements dictated by the Nyquist sampling theorem. Moreover, using one-third of the measurements or less, the new compressive sampling method can provide over 20 dB better denoising capability than oversampling with classical Fourier theory.  
\end{abstract}


\maketitle




\section{Introduction}
Band-limited Spherical Wavefunction (SW) expansions in 3D and their restrictions to a sphere, Spherical Harmonic (SH) expansions, have become a key tool in many acoustics applications. Recent high-interest applications of these series expansions range from surround sound~\cite{polettiThreeDimensionalSurroundSound2005,ben-hurLoudnessStabilityBinaural2019,zuoIntensityBasedSpatial2020}, spherical acoustic holography~\cite{zotterSamplingStrategiesAcoustic2009,jacobsenFieldAcousticHolography2011,fernandez-grandeCompressiveAcousticHolography2018}, and acoustic levitation~\cite{anderssonAcousticLevitationSuperposition2019,zehnterAcousticLevitationMie2021} to beam-forming/source localization~\cite{parkSoundfieldAnalysisPlanewave2005,fernandez-grandeCompressiveSensingSpherical2016}, directivity characterization~\cite{ahrensComputationSphericalHarmonics2020,ahrensComputationSphericalHarmonic2021}, ultrasonic medical imaging~\cite{tutarSemiautomatic3DProstate2006,velasquez-rodriguezAutomaticSegmentationFetal2015} and material characterization~\cite{lanSphericalHarmonicApproach2015,lanDirectVolumetricMeasurement2018}, and even electromagnetic (EM) applications like spherical near-field to far-field transformations (SNF2FFTs)~\cite{hansenSphericalNearfieldAntenna1988}. The utility of the SW and SH expansions are quite broad. In surround sound settings~\cite{polettiThreeDimensionalSurroundSound2005,ben-hurLoudnessStabilityBinaural2019,zuoIntensityBasedSpatial2020}, spherical acoustic holography~\cite{zotterSamplingStrategiesAcoustic2009,jacobsenFieldAcousticHolography2011,fernandez-grandeCompressiveAcousticHolography2018}, and acoustic or EM directivity characterizations~\cite{ahrensComputationSphericalHarmonics2020,ahrensComputationSphericalHarmonic2021,hansenSphericalNearfieldAntenna1988,loschenbrandFastAntennaCharacterization2016,fuchsFastAntennaFarField2017,culotta-lopezCompressedSamplingSpherical2018,bangunCoherenceBoundsSensing2018,bangunSensingMatrixDesign2020}, one needs explicit knowledge of the SW or SH series coefficients to reconstruct or reproduce a given sound or EM field. A similar situation is present in ultrasonic medical imaging~\cite{tutarSemiautomatic3DProstate2006,velasquez-rodriguezAutomaticSegmentationFetal2015}, where SH series coefficients are needed to model the shape of various organs inside of the body. The usage of SWs and SHs in crystallography is slightly different; in this application the SW or SH coefficients are used to relate an easily measurable quantity to the crystalline texture of polycrystalline material~\cite{lanSphericalHarmonicApproach2015,lanDirectVolumetricMeasurement2018}. Moreover, in acoustic levitation, ultrasound, and even room transfer function estimation~\cite{samarasingheSphericalHarmonicsBased2018}, the need to know the acoustic field of a device to a high level of detail presents a future application for SW/SH characterizations, in particular acoustic SNF2FFTs~\cite{wittmannProbecorrectedSphericalNearfield1992}. Such characterizations are especially important for spherical arrays that are being used to generate or characterize sound fields because minor transducer errors in the array can lead to major performance degradation in certain scenarios~\cite{raoEffectTransducerMismatch2014}.

Estimating the SW or SH expansion coefficients of an acoustic field requires first taking measurements on a fixed radius sphere using a spherical array of microphones~\cite{polettiThreeDimensionalSurroundSound2005,zuoIntensityBasedSpatial2020} or, in the more general case, higher-order probes~\cite{wittmannProbecorrectedSphericalNearfield1992}, i.e., extended geometry probes sensitive to high order SH/SW modes ($m>1$ modes). From these measurements, one can then use integral approaches~\cite{polettiThreeDimensionalSurroundSound2005,lanSphericalHarmonicApproach2015,lanDirectVolumetricMeasurement2018} or, as has become quite common, a linear inverse problem~\cite{polettiThreeDimensionalSurroundSound2005,ben-hurLoudnessStabilityBinaural2019,zuoIntensityBasedSpatial2020} to solve for the SW coefficients. According to the Nyquist sampling theorem, the number of measurements $M$ required to accurately estimate a field's coefficients in the band-limited SW/SH series scales with the square of the band-limit~\cite{wittmannProbecorrectedSphericalNearfield1992,polettiThreeDimensionalSurroundSound2005} (here band-limit refers to the highest degree SH/SW needed to describe the spatial distribution of the field). Depending on whether a classical integration or linear inverse problem is used, the constant scaling coefficients of this quadratic relationship can vary~\cite{polettiThreeDimensionalSurroundSound2005,lanSphericalHarmonicApproach2015}. When using a measurement probe sensitive to only $m\leq 1$ modes, this also holds for vector SWs/SHs in EM applications~\cite{hansenSphericalNearfieldAntenna1988}. For even small band-limits, the number of measurements can be time-consuming and turn into hundreds or even thousands of measurements~\cite{polettiThreeDimensionalSurroundSound2005,lanSphericalHarmonicApproach2015}. Thus, characterizing an acoustic field can require many microphones in a spherical array (e.g., in spherical holography, source localization/beam-forming, directivity characterization, etc.)\ or be time consuming for using if re-positioning the experimental apparatus (e.g., the source/receiver microphone/higher-order probe or a polycrystalline material). As a note, in EM SNF2FFTs, measurement numbers fare even worse since typical devices can require hundreds of thousands of measurements due to their band-limit and noise requirements~\cite{loschenbrandFastAntennaCharacterization2016, fuchsFastAntennaFarField2017}. 

In many of the above-mentioned applications, methods relating to sparse signal processing have been of interest in order to speed up measurement times, reduce the need for many measurement devices, or decrease the number of transducers needed to reproduce a sound field~\cite{lilisSoundFieldReproduction2010,chardonNearfieldAcousticHolography2012,jinTheoryDesignMultizone2015, fernandez-grandeCompressiveAcousticHolography2018,ben-hurLoudnessStabilityBinaural2019}. When the sound fields of interest satisfy certain properties, e.g., the sound field is from a symmetric loudspeaker or the field has a smooth, regular, or symmetric spatial distribution, the SW/SH coefficients can be considered as sparse or compressible (i.e., approximately sparse). Here, sparse means that the coefficients contain mostly zero entries with few non-zero values. In these cases where the coefficients are sparse or compressible, Compressive Sampling (CS) can be used to accurately solve the linear inverse problem for the SW/SH coefficients while requiring fewer measurements than needed in integral approaches or to make the linear inverse problem fully determined~\cite{candesDecodingLinearProgramming2005,candesRobustUncertaintyPrinciples2006,candesNearOptimalSignalRecovery2006,rauhutRandomSamplingSparse2007, rudelsonSparseReconstructionFourier2008,rauhutSparseLegendreExpansions2012,foucartMathematicalIntroductionCompressive2013}. Such a reduction in the required number of measurements can allow for reduced measurement times and, for microphone arrays, require fewer microphones to be used. 

\subsection{Contributions and Relation to Other Work}

To derive a CS approach for the many acoustics applications described above as well as EM applications like SNF2FFTs, we prove a CS guarantee for a series of Wigner $D$-functions. SH or SW function series are special cases of Wigner $D$-function series, and so the approach developed also specializes to these two cases. Specifically, in the case where measurements are performed by moving an ``ideal'' microphone (measuring a perfect point of the field) to different measurement positions or using a spherical array of ``ideal'' microphones, the Wigner $D$-function expansion reduces to the SH (thus SW) expansion by using only a certain subset of Wigner $D$-functions. Alternatively, in the case where measurements are taken using an extended non-ideal probe, the Wigner $D$-function series is required unless simplifying assumptions are made, e.g., using an ideal measurement probe. This is a direct result of accounting for the probe's sensitivity to SWs in its coordinate frame and carrying out the appropriate transformations to relate the measurements taken and the SW expansion coefficients of interest~\cite{wittmannProbecorrectedSphericalNearfield1992}. For mathematical context, the Wigner $D$-functions are an irreducible representation of the symmetry group of the sphere, the rotation group $\sot$, and they form an orthogonal basis for band-limited functions on $\sot$~\cite{varshalovichQuantumTheoryAngular1988}.

For a particular linear inverse problem, the number of measurements required and where these measurements should be taken are key factors in determining the success of CS. One approach is to use experimentation or to algorithmically find measurement positions that minimize the coherence of the measurement matrix (which depends on the number of measurements and their positions) in the inverse problem while remaining within constraints dictated by device properties, e.g., microphone radii, positioning accuracy, etc. For applications using a Wigner $D$-function series to solve for its series coefficients, experimentally tested approaches have been given considerable study~\cite{corneliusCompressedSensingApplied2016,fuchsFastAntennaFarField2017,fuchsCompressiveSensingApproach2018} and coherence based analyses have also been of interest~\cite{culotta-lopezCompressedSamplingSpherical2018,bangunCoherenceBoundsSensing2018,bangunSensingMatrixDesign2020}. However, such approaches do not establish the required number of measurements, guarantee robustness to noise, guarantee robustness to small increases in signal sparsity, or guarantee robustness to small decreases in measurement number.

Alternatively, the drawbacks of experimentation or coherence analyses can be avoided if the measurement matrix satisfies certain properties. Conventionally, the Robust Nullspace Property (RNP) guarantees methods like Quadratically Constrained Basis Pursuit (QCBP) produce accurate solutions for the unknown signal. A sufficient condition that provides strong guarantees for satisfying the RNP is a measurement matrix that satisfies the Restricted Isometry Property (RIP). Verifying the RIP for a given measurement matrix is NP-hard~\cite{tillmannComputationalComplexityRestricted2014}. Due to this fact, probabilistic approaches showing that a measurement matrix satisfies the RIP with high probability are normally used~\cite{rauhutSparseLegendreExpansions2012, bangunSensingMatrixDesign2020, valdezRestrictedDomainCompressive2022}. Such approaches have been used to give RIP-based results for Wigner $D$-functions~\cite{bangunSensingMatrixDesign2020, valdezRestrictedDomainCompressive2022}. The drawback of the existing probabilistic RIP-based guarantees, however, is that they require sampling at arbitrary random positions on $\sot$ to get robust theoretical guarantees. This requirement can be difficult or impossible for most measurement systems or microphone arrays; arbitrary points can be too close physically in multi-probe systems and highly time-inefficient in single probe systems. This situation contrasts the coherence-based methods like those in~\cite{culotta-lopezCompressedSamplingSpherical2018,bangunCoherenceBoundsSensing2018,bangunSensingMatrixDesign2020}, where hard/impossible measurement patterns can be deliberately excluded from use. 

In this article, we develop a CS method to estimate Wigner $D$-function series expansion coefficients where measurements are taken randomly {\em from a fixed grid}. Moreover, this CS method has robust theoretical guarantees specifying the number of measurements needed and bounding the error of the estimated coefficients. Thus, our work negates the problematic need for sampling at arbitrary positions in space as present in existing approaches~\cite{rauhutSparseLegendreExpansions2012, bangunSensingMatrixDesign2020, valdezRestrictedDomainCompressive2022} and gives stronger and more prescriptive theoretical guarantees than coherence based analyses. Thus, we show that CS guarantees can be applied to the linear inverse problem arising from solving for the Wigner $D$-function coefficients without arbitrarily positioned measurements on $\sot$. Additionally, our approach requires fewer measurements than the classical Nyquist sampling theorem requires. To the best of our knowledge, this is the first result giving theoretical guarantees for CS recovery of the coefficients in a series of Wigner $D$-functions where measurements are selected from a fixed grid on $\sot$ (sphere in the SW/SH special cases).

\subsection{Outline of Results}
In the present article, we provide a method to apply CS to recover the coefficients of a series of Wigner $D$-functions, and thus SW or SW series, that only requires selecting samples from a pre-defined grid on $\sot$ (sphere in the special case of SWs/SHs). In this problem we assume that the field measured, $w(\alpha,\beta,\gamma)$, can be written as a band-limited series of the Wigner $D$-functions. The arguments $\alpha$, $\beta$, and $\gamma$ parametrize a point in $\sot$ corresponding to a physical position where the field can be measured. In this setup, $\alpha$ and $\gamma$ are $2\pi$ periodic while $\beta$ is typically taken to be in $[0,\pi]$. It is well known, however, that the Wigner $D$-functions naturally possess a periodicity when taking $\beta \in [0, 2 \pi )$~\cite{varshalovichQuantumTheoryAngular1988}. Moreover, this periodicity is captured in a well-known Fourier expansion for the Wigner $D$-functions~\cite{hansenSphericalNearfieldAntenna1988}. Thus, the approach we take to solve for the coefficients of a Wigner $D$-function series is to utilize this natural domain extension, periodicity, and Fourier expansion. Since we assume $w(\alpha,\beta,\gamma)$ is a series of Wigner $D$-functions, the periodic extension of the Wigner $D$-function naturally induces one in $w$ and makes $w$ periodic in all of its arguments by letting $\beta$ be in $[0,2\pi)$. This domain extension takes the function $w$ on $\sot$ and maps it to the 3-torus $\tthre$, which is a double cover of $\sot$. 

Now considering the periodically-extended Wigner $D$-functions and $w(\alpha,\beta,\gamma)$, we transform the problem to the Fourier basis. This allows us to treat the solution of the Wigner $D$-function inverse problem as a multi-dimensional Discrete Fourier Transform (DFT) problem. Moreover, this transformation is carried out in a way that preserves important sparsity structures in $w$. In this multi-dimensional DFT form, we can then sample a subset of the positions to achieve compressive sampling for the band-limited series of Wigner $D$-functions (see Theorem \ref{thm:subsampled_3ddft}). 

In a bit more mathematical detail, we start with the problem of solving for the vector $a$ from the inverse problem, 
\begin{equation}\label{eq:wigner_D_formulation}
    w = \Phi_D a +\eta,
\end{equation}
where $w$ is the vector of measurements of $w(\alpha,\beta,\gamma)$ at a set of points $(\alpha_j,\beta_j,\gamma_j)\in\sot$, $\Phi_D$ is the measurement matrix whose rows contain the Wigner $D$-functions $\WD{n}{\mu}{m}(\alpha_j,\beta_j,\gamma_j)$, $a$ is the vector of coefficients in the Wigner $D$-function series for $w(\alpha,\beta,\gamma)$, and $\eta$ is additive measurement noise. Here, we have standard ranges for the arguments of the Wigner $D$-function, $\alpha\in[0,2\pi)$, $\beta\in[0,\pi]$, and $\gamma\in[0,2\pi)$. Note, the Wigner $D$-functions relate to the spherical harmonics in a form like $\WD{n}{\mu}{0}(\alpha,\beta,\gamma) = \t{c}{_n^\mu}\t{Y}{_n^{-\mu}}(\beta,\alpha)$ or $\WD{n}{0}{m}(\alpha,\beta,\gamma) = \t{c}{_n^m}\t{Y}{_n^{-m}}(\beta,\gamma)$, where the $c$ coefficients are constants depending on its indices~\cite{varshalovichQuantumTheoryAngular1988}. Thus, SW/SH series can be considered a special case of \eqref{eq:wigner_D_formulation}.

Without making any further assumptions beyond $w(\alpha, \beta,\gamma)$ being representable as a series of Wigner $D$-functions, we recognize that if we extend $\beta$ to be in $[0,2\pi)$, the Wigner $D$-functions in $\Phi_D$ become periodic in all three arguments $(\alpha,\beta,\gamma)$~\cite{varshalovichQuantumTheoryAngular1988}. With this periodic domain extension, instead of $(\alpha,\beta,\gamma)$ being on $\sot$, they are taken to be on $\tthre$. Since $w(\alpha,\beta,\gamma)$ is the Wigner $D$-function series with coefficients in $a$, $w(\alpha,\beta,\gamma)$ also becomes periodic. In terms of measurements, this amounts to letting the polar angle of measurements wrap completely around the sphere on which measurements are taken. For a spherical microphone array, this is a reinterpretation of the existing microphone positions. Each microphone would have one position with $0 \leq \beta\leq \pi$ and a second position with $\pi < \beta < 2\pi$. As discussed later, this means the measurements from a specific microphone may be used twice. Additionally, for a high-order probe like those in~\cite{hansenSphericalNearfieldAntenna1988,wittmannProbecorrectedSphericalNearfield1992}, the domain extension means the second rotation in the set of Euler rotations in the $zy'z'$ convention is extended to a full $2\pi$ range. With these physical pictures in mind, the periodicity of $w(\alpha,\beta,\gamma)$ with an extension in $\beta$ becomes more intuitive. Note, the periodic extension of the domain and resulting re-use of measurements can appear at first glance to not benefit the prospects of CS. In particular, prospects would not improve because it would merely add repeated rows to the measurement matrix in \eqref{eq:wigner_D_formulation}. However, we do not argue to use the Wigner $D$-function domain problem as is; rather, we propose transforming the problem to the spatial Fourier domain. In this alternative formulation, these physically identical measurements will constitute distinct rows of a new measurement matrix.

Due to the periodicity of the Wigner $D$-functions, there is a transformation, which we denote with $B$, that takes the problem from a Wigner $D$-function basis with arguments $\alpha,\;\beta,\;\gamma\in[0,2\pi)$ to the Fourier series basis. That is, we can write the problem as 
\begin{align}
    w &= \Phi_F b +\eta, \\
    b &= B a,
\end{align}
where $\Phi_D=\Phi_F B$. Here $\Phi_F$ is the measurement matrix whose rows contain the basis functions for the three-dimensional Fourier series, and $b$ is the vector containing the Fourier coefficients for $w$ on $\tthre$. Fortunately, the $B$ matrix derives from the well-known Fourier expansion for the Wigner $D$-functions and can be computed directly from the Wigner $D$-functions. Importantly, it is the case that $B$ is well-conditioned and increases the sparsity level of the problem in a bounded way in situations where the sparsity comes from the field coefficients using a few $m,\;\mu$ subspaces or only low-frequency functions in the Wigner $D$-function basis. Thus, to solve \eqref{eq:wigner_D_formulation}, we can first solve for the Fourier coefficients $b$, and then solve for the Wigner $D$-function (or SW/SH) coefficients $a$. When one is simply interested in the special cases of SW or SH series, this directly gives the coefficients desired. For high-order probes (e.g., in SNF2FFTs), $a$ can be used to calculate the SW coefficients of the speaker/emitter after factoring out the appropriate transformation information~\cite{wittmannProbecorrectedSphericalNearfield1992,hansenSphericalNearfieldAntenna1988}.

In the above, we transformed the problem in \eqref{eq:wigner_D_formulation} from a Wigner $D$-function series on $\sot$ to a Fourier problem on $\tthre$. If we suppose that measurements are taken at a selection of points from an equiangular grid covering $\tthre$ at the Nyquist rate (which is also a Nyquist sampling on $\sot$), then $\Phi_F$ becomes a sub-sampled three-dimensional DFT (3DDFT) matrix. We denote the 3DDFT matrix as $U_F$. Since the measurements form a sub-sampled 3DDFT, we can write the problem as
\begin{equation}
    w = P_\Omega U_F b + \eta, 
\end{equation}
where $P_\Omega$ is the matrix selecting a subset of $M$ rows from $U_F$. With the problem in \eqref{eq:wigner_D_formulation} cast in this way, we can apply standard CS recovery guarantees for sub-sampled unitary measurement matrices~\cite{foucartMathematicalIntroductionCompressive2013}. Thus, we achieve compressive measurements from a sub-selection of measurements from a pre-defined grid on $\tthre$ (and so $\sot$). To get these robust reconstruction guarantees from CS, the number of measurements, $M$, must scale as (see Foucart and Rauhut Corollary 12.38~\cite{foucartMathematicalIntroductionCompressive2013} and our main result, \Cref{thm:subsampled_3ddft})
\begin{equation}
    M\geq \til{C} s_F \ln^4(N_F),
\end{equation} 
where $N_F$ is the size of the band-limited three-dimensional Fourier basis, $s_F$ is the sparsity in this basis, and $\til{C}$ is a constant. Transforming this equation for $M$ into a form depending only on Wigner $D$-function basis information, one gets a \tit{worst-case} scaling of (see \Cref{thm:subsampled_3ddft} and Remark~\ref{rem:wigner_d_sparsity_vs_fourier})
\begin{equation}
    M\geq \til{C}' N_D^{1/3} s_D \ln^4(N_D),
\end{equation}
where $\til{C}'$ is a constant, $s_D$ is the sparsity in the Wigner $D$-function basis, and $N_D$ is the size of the Wigner $D$-function basis.

\subsection{Structure of this Paper}
The remainder of this paper is structured as follows. Section \ref{sec:notation} provides the notation used throughout the paper. \Cref{sec:background} contains the background information on the Wigner $D$-functions and field measurements in the general case of using high-order probes in~\Cref{sec:near-field_measurments} along with the results we need from the CS literature in \Cref{sec:cs_preliminaries}. Section \ref{sec:fourier_basis_cs} then gives the transformation of the inverse problem in \eqref{eq:wigner_D_formulation} from the Wigner $D$-function formulation to the Fourier formulation as well as the CS guarantees for this problem with the gridded sampling of $\tthre$. We follow with numerical examples in \Cref{sec:numerical_examples}. This section includes investigations into the effect that transforming from the Wigner $D$-function formulation to the Fourier formulation has on the sparsity of the coefficient vector, \Cref{sec:rand_sparsity,sec:speaker_sparsity}, and then examples of CS recovery in the Fourier formulation, \Cref{sec:cs_recovery_numerics}. Lastly, we provide a conclusion in \Cref{sec:conclusion}.

\subsection{Notation} \label{sec:notation}
Throughout this paper, we use the following notation and conventions. The sum $\sum_{m,\mu,m'=-n}^{n}$ is used to mean $\sum_{m=-n}^{n} \sum_{\mu=-n}^{n} \sum_{m'=-n}^{n}$. We use $\mi=\sqrt{-1}$ as the unit imaginary number. An over-line represents complex conjugation, e.g., $\overline{a}$. We represent the Hermitian conjugate of a vector or matrix with $^H$, e.g., $a^H = \overline{a^T}$, where $T$ denotes the transpose operation. The norm $\|\cdot\|_2$ is the standard $\ell_2$ vector norm. $\|\cdot\|_\infty$ is either the $\ell_\infty$ or $L_\infty$ norm, which should be discernible from context. As usual, for a vector, $\|a\|_\infty = \max_i(|a_i|)$, and for a function, $\|f\|_\infty = \inf\{c\geq0:|f(x)|\leq c\; \text{for almost every} \; x\}$. For Euler rotations, we use the $zy'z'$ and passive transformation conventions~\cite{varshalovichQuantumTheoryAngular1988}. We use i.i.d.\ to abbreviate independently and identically distributed.

\section{Background} \label{sec:background}
In this section, we develop the background for our work in the context of spherical field measurements with general probes and also state the requisite CS reconstruction guarantees from the literature.

\subsection{Spherical Field Measurements and Compressive Sampling} \label{sec:near-field_measurments}
When measuring acoustic fields, it is often assumed that the transducers used are omnidirectional point receivers that can measure the field of interest directly at their position, see for example~\cite{polettiThreeDimensionalSurroundSound2005}. However, when high-order transducers or arrays are used for measurement, there will be inherent directional dependence on the directivity of the field probe. To account for this property in spherical field measurements, one must include the directionality of the measurement device in the calculation of the field from measurements. The directionality of the field probe can be defined in terms of its sensitivity to SWs, called the receiving coefficients and denoted by $\t{R}{_\nu^\mu}$. The inclusion of this directionality is called probe correction and is discussed in detail in~\cite{wittmannProbecorrectedSphericalNearfield1992}. 

\begin{figure}[ht]
    \centering
    \includegraphics[width=\reprintcolumnwidth]{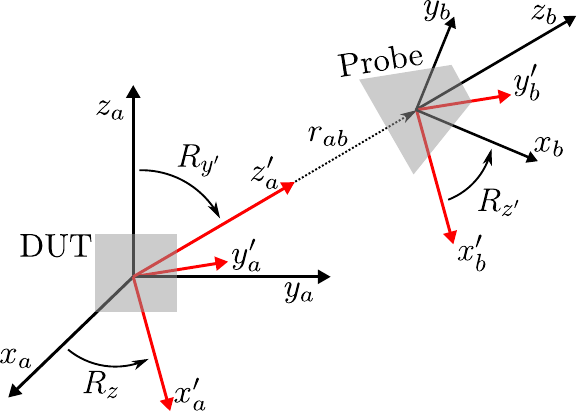}
    \caption{\tit{Transformation to Probe Coordinates.} The measurement position of a probe is specified by three Euler angles $(R_z, R_{y'}, R_{z'})$ and a fixed distance to the spherical scanning surface, $r_{ab}$. The device under test (DUT) is centered on coordinate system $(x_a, y_a, z_a)$ and the probe on coordinate system $(x_b, y_b, z_b)$. To transform from the DUT to the probe one first rotates by $R_z$ about $z_a$, giving $(x_a^1, y_a^1, z_a^1)$. Then one rotates by $R_{y'}$ about $y_a^1$ bringing the coordinate system into $(x_a', y_a', z_a')$. Next $(x_a', y_a', z_a')$ is translated by $r_{ab}$ to $(x_b', y_b', z_b')$ and a final rotation of $R_{z'}$ about $ z_b'$ brings one into $(x_b, y_b, z_b)$.}
    \label{fig:rotation_diagram}
\end{figure}

In summary, probe correction for spherical field measurements is carried out as follows. First one considers the field of interest as a band-limited series of outgoing spherical waves (if a radiator is contained in the sphere of measurements) or standing waves (if no source is in the sphere of measurements). We denote this band limit as $n_{\max}$. As a note, if a radiator is present, $n_{\max}$ is related to the physical size of the radiator. In the no radiator case, $n_{\max}$ relates to the frequency content in the field of interest~\cite{polettiThreeDimensionalSurroundSound2005}. A coordinate transformation is then performed so that the probe-centered coordinate system, which lies on the surface of the sphere enclosing the radiator, properly relates measurements to the coordinate system centered on the radiator. This is done using a rotation and  translation. The rotation can be specified as a set of Euler rotations parametrized in the $zy'z'$ and passive transformation conventions. The resulting expression for the measured field is
\begin{equation}\label{eq:wigner_D_expanison}
    \begin{split}
        w_j & = w(\alpha_j,\beta_j,\gamma_j)  \\
        & = \sum_{n=0}^{n_\text{max}}\sum_{m=-n}^n \sum_{\mu=-n}^n\acoef{n}{m\mu} \WD{n}{\mu}{m}(\alpha_j,\beta_j,\gamma_j) + \eta_j,
    \end{split}
\end{equation}
where $\eta_j$ are elements of the additive measurement noise $\eta$. Here $j$ indexes arguments of the Wigner $D$-function,  $(\alpha_j,\beta_j,\gamma_j) \in \sot$, corresponding to the $j$th measurement position. These relationship between the coordinates in $\sot$ and the Euler angles of measurement point $j$, $(R_z, R_{y'}, R_{z'})$, is given by $(\alpha_j,\beta_j,\gamma_j) = (-R_{z'}, -R_{y'}, -R_z)$ or equivalently $(\alpha_j,\beta_j,\gamma_j) = (\pi-R_{z'}, R_{y'}, \pi-R_z)$~\cite{varshalovichQuantumTheoryAngular1988}. This is depicted in \Cref{fig:rotation_diagram}. 
Additionally, $\WD{n}{\mu}{m}(\alpha,\beta,\gamma)$ is the Wigner $D$-function and its indices satisfy $n\in\{0,1,\cdots, n_\text{max}\}$ and $m,\mu \in\{-n,-n+1,\cdots,n-1,n\}$. The total number of Wigner $D$-functions and thus coefficients $N_D = (n_\text{max}+1)(2n_\text{max}+1)(2n_\text{max}+3)/3$. The Wigner $D$-function is defined as
\begin{equation}\label{eq:wigner_D}
    \t{D}{_n^{\mu m}}\left(\alpha,\beta,\gamma\right) = e^{-\mi \mu \alpha}\t{d}{_n^{\mu m}}(\beta) e^{-\mi m \gamma},
\end{equation}
where $\t{d}{_n^{\mu m}}$ is the purely real Wigner $d$-function defined by~\cite{varshalovichQuantumTheoryAngular1988} (Section 4.3.1 eq. 4)
\begin{equation}\label{eq:wigner_d}
    \begin{split}
        \t{d}{_n^{\mu m}}(\beta) =\; & (-1)^{\mu -m} \\
        & \times \sqrt{(n+m)!(n-m)!(n+\mu)!(n-\mu)!} \\
    & \times \sum_{\sigma = \max \left(0, m-\mu\right)}^{\min\left(n+m,n-\mu\right)} \xi_\sigma, \\
    \xi_\sigma =\;&\frac{
        (-1)^\sigma \left(
            \cos\frac{\beta}{2}\right)
        ^{2n-2\sigma+m-\mu} \left(
            \sin\frac{\beta}{2}
        \right)^{2\sigma-m+\mu}
    }{
        \sigma! (n+m-\sigma)!(n-\mu-\sigma)!(\mu-m+\sigma)!
    }.
    \end{split}
\end{equation}
The Wigner $d$-function's indices satisfy the same restrictions as those of the $D$-function. The spherical harmonics are a special case of the the Wigner $D$-functions, given by ~\cite{varshalovichQuantumTheoryAngular1988} (Section 4.17 eq. 1),
\begin{align}
    \t{Y}{_n^\mu}(\beta, \alpha)& = (-1)^m \frac{\sqrt{2n + 1}}{4\pi} \WD{n}{-\mu}{0}(\alpha, \beta,\gamma) \\
    \t{Y}{_n^m}(\beta, \gamma)& = \frac{\sqrt{2n + 1}}{4\pi} \WD{n}{0}{-m}(\alpha,\beta,\gamma).
\end{align}

In \eqref{eq:wigner_D_expanison} the coefficients $\acoef{n}{m\mu}$ contain the original SW coefficients, the translation transformation information, and the probe receiving coefficients. In particular, the $\acoef{n}{m\mu}$ take a form like $\acoef{n}{m\mu} = \Acoef{n}{m} \sum_\nu \t{S}{_{n\nu}^{\mu \mu}}(r_{ab}) \t{R}{_\nu^\mu}$, where $\Acoef{n}{m}$ are the SW coefficients, and $\t{S}{_{n\nu}^{\mu \mu}}$ is the translation operator for SWs with a translation distance of $r_{ab}$. The explicit form of $\t{S}{_{n\nu}^{\mu \mu}}$ depends on whether the original field expansion is outgoing or standing waves~\cite{martinMultipleScatteringInteraction2006}.

The paragraphs above setup a linear inverse problem we wish to solve using CS. However, for CS to be a valid approach we must establish cases where coefficient sparsity is a valid assumption. In cases where the field of interest is highly symmetric with respect to the radiation coordinate system, there will be few non-zero $\Acoef{n}{m}$, so regardless of probe type the $\acoef{n}{m\mu}$ will be sparse/compressible~\cite{culotta-lopezRadiationCenterEstimation2017,culotta-lopezFastNearfieldAntenna2021}. In a more specific case, if the probe is small, i.e., $k_\lambda r_\mrm{probe}$ is small (where $k_\lambda$ is the wavenumber of the field of interest and $r_\mrm{probe}$ is the smallest sphere circumscribing the probe) or highly rotationally symmetric, the $\t{R}{_{\nu}^{\mu}}$ will be approximately zero for $\mu \geq |1|$ and larger $\nu$, so again the $\acoef{n}{m\mu}$ will be sparse/compressible even if the $\Acoef{n}{m}$ are not totally sparse themselves. In contrast, in cases where an omnidirectional receiver that directly measures the field of interest is used, \eqref{eq:wigner_D_expanison} is equal to the field sampled at the position $j$. This can be interpreted as the translation factors multiplied by the Wigner $D$-functions collapsing down to be the SW/SH expansion for the field at the point $j$~\cite{wittmannProbecorrectedSphericalNearfield1992}. Due to this collapsing of the series expansion, the sparsity assumption must hold for the SW/SH coefficients $\Acoef{n}{m}$ directly. Note, our method is general and does not depend on the probe selection, only the validity of the sparsity assumption does.

For comparison with our method in Section~\ref{sec:fourier_basis_cs}, we now give the existing scaling in the number of measurements required to give successful and robust CS recovery for Wigner $D$-functions and SHs. Note, both results require random measurements on arbitrary positions of their domain. For Wigner $D$-functions the number of such measurements must satisfy (see Theorem 3 from Bangun, et al.~\cite{bangunSensingMatrixDesign2020})
\begin{equation}\label{eq:wigner_d_m_scaling}
    M\geq \tilde{C}' N_D^{1/6}s_D\ln^4(N_D),
\end{equation}
where $s_D$ is the sparsity in the Wigner $D$-function basis. Here $\til{C}'\geq0$ is constant. For the case of the sphere, i.e., SW/SH expansion, $M$ has the same form of scaling but with $N_D$ replaced with the number of band-limited SHs, $N_{SH}=(n_{\max}+1)^2$, and $s_D$ replaced with the sparsity in the SH basis, $s_{SH}$~\cite{burqWeightedEigenfunctionEstimates2012}, i.e.,
\begin{equation}\label{eq:sh_scaling}
    M\geq \tilde{C}' N_{SH}^{1/6}s_{SH}\ln^4(N_{SH}).
\end{equation}

Though the CS guarantees mentioned above are useful in establishing theoretical viability and requirements for using CS to solve \eqref{eq:wigner_D_expanison} for its coefficients, their requisite random sampling is problematic. As mentioned in the introduction, this is because many measurement systems struggle to reach arbitrary points on $\sot$ or the sphere, and furthermore, random points on $\sot$ or the sphere can be too close together for some measurement setups and even time-consuming when compared to regular patterns. Thus, these CS results provide mixed gains. On one hand, they reduce the required number of measurements. On the other hand, they can give impossible measurement positions (when they are too close for arrays) or increase measurement times due to random positioning requirements. In contrast, coherence-based analyses like those in \cite{culotta-lopezCompressedSamplingSpherical2018,bangunCoherenceBoundsSensing2018, bangunSensingMatrixDesign2020} provide regular patterns for measurements that can be carried out rapidly but these do not give strong theoretical guarantees in the sense of prescribing a needed number of measurements for QCBP to be successful with a guaranteed accuracy and robustness to noise/small decreases in measurement number. As we demonstrate in \Cref{sec:fourier_basis_cs}, it is possible to maintain theoretical recovery guarantees while using more regular sampling patterns on $\sot$ or the sphere.

\subsection{Compressive Sampling Preliminaries}\label{sec:cs_preliminaries}
For our results that follow, we will need the following definitions and results from the CS literature. 

\begin{definition}[Best $s$-Sparse Approximation Error~{\cite[p.~42, Def.~2.2]{foucartMathematicalIntroductionCompressive2013}}]
    Given a vector $x\in\mathbb{C}^N$, the best $s$-sparse approximation error in the $\ell_p$ norm is
    \begin{equation}
        \sigma_s(x)_p = \min_{z\in\mathbb{C}^N:\|z\|_0\leq s} \|z-x\|_p.
    \end{equation}
\end{definition}

\begin{definition}[Restricted Isometry Property (RIP)~{~\cite{bangunSensingMatrixDesign2020}, \cite[p.~133, def.~6.1]{foucartMathematicalIntroductionCompressive2013}}]
    A matrix $\Phi\in \mathbb{C}^{M\times N}$ satisfies the restricted isometry property of order $s$ with constant $\delta \in (0,1)$ if the following holds for all $s$-sparse vectors $x\in \mathbb{C}^N$
    \begin{equation}
        (1-\delta) \|x\|_2^2\leq \|\Phi x\|_2^2\leq (1+\delta) \|x\|_2^2,
    \end{equation}
    where the smallest $\delta$ denoted by $\delta_s$ is called the restricted isometry constant of order $s$.
\end{definition}

\begin{theorem}[RIP for Bounded Orthonormal Systems (BOS)~{\cite[p.~405, Thm.~12.31]{foucartMathematicalIntroductionCompressive2013}}] \label{thm:rip_bos}
    Consider a set  of bounded orthonormal basis functions $\phi_i:\mcl{D}\rightarrow \mathbb{C},\; i\in\{1,2,\cdots,N\}$ that are orthonormal with respect to a probability measure $\rho$ on the measurable space $\mcl{D}$. Consider the matrix $\Phi\in \mathbb{C}^{M \times N}$ with entries 
    \begin{equation}
        \phi_{ji} = \phi_i(t_j),\; j\in\{1,2,\cdots,M\},\; i\in\{1,2,\cdots,N\}
    \end{equation}
    constructed with i.i.d.\ samples of $t_j$ from the measure $\rho$ on $\mcl{D}$. Suppose the orthonormal functions are bounded such that $\sup_{i\in\{1,\cdots,N\}} \|\phi_i\|_\infty\leq K$. If
    \begin{equation}
        M \geq C_0 \delta^{-2} K^2 s \ln^4(N)
    \end{equation}
    then with probability at least $1-N^{-\ln^3(N)}$, the restricted isometry constant $\delta_s$ of $\frac{1}{\sqrt{M}}\Phi$ satisfies $\delta_s\leq \delta$ for $\delta \in(0,1)$. The constant $C_0\geq0$ is universal.
\end{theorem}

\begin{corollary}[RIP For Unitary Matrices (BOS)~{\cite[p.~405, Thm.~12.31 \& p.~405, Cor.~12.38]{foucartMathematicalIntroductionCompressive2013}}] \label{cor:unitary_rip}
    Let $U\in\mbb{C}^{N\times N}$ be a unitary matrix with entries bounded from above by $K/\sqrt{N}$. Let $\Phi\in\mbb{C}^{M\times N}$ be the sub-matrix of $U$ acquired by selecting a subset of rows of size $M$ from $U$ uniformly at random among all subsets of size $M$. If 
    \begin{equation}
        M \geq C_0 \delta^{-2} K^2 s \ln^4(N)
    \end{equation}
    then with probability at least $1-N^{-\ln^3(N)}$, the restricted isometry constant $\delta_s$ of $\frac{1}{\sqrt{M}}\Phi$ satisfies $\delta_s\leq \delta$ for $\delta \in(0,1)$. The constant $C_0\geq0$ is universal.
\end{corollary}

\begin{theorem}[Sparse Recovery for RIP Matrices~{\cite[p.~144, Thm.~6.12]{foucartMathematicalIntroductionCompressive2013}}] \label{thm:sparse_recovery_rip}
    Suppose that the matrix $\Phi\in\mathbb{C}^{M\times N}$ has restricted isometry constant $\delta_{2s} < 4/\sqrt{41} \approx 0.6246$. Suppose that the measurements are taken with $\Phi$ and are noisy, $y=\Phi x +\eta$, with $\|\eta\|_\infty\leq \epsilon$. If $\widehat{x}$ is the solution to 
    \begin{equation}
        \widehat{x} = \arg \min_{z\in\mathbb{C}^N}  \|z\|_1 \; \text{subject to} \; \|y-\Phi z\|_2 \leq \sqrt{M}\epsilon
    \end{equation}
    then 
    \begin{equation}
        \|x-\widehat{x}\|_2 \leq C_1\left( \frac{\sigma_s(x)_1}{\sqrt{s}} + \epsilon \right),
    \end{equation}
    where $C_1\geq0$ only depends on $\delta_{2s}$.
\end{theorem}

\section{On-Grid Compressive Sampling for Spherical Field Measurements} \label{sec:fourier_basis_cs}

\begin{lemma}[Solving for Wigner $D$-function Series Coefficients in the Discrete Fourier Basis] \label{lem:dft_cs_for_wD_series}
    Let $w(\alpha, \beta, \gamma)\in L_2(\sot)$ have a band-limit $n_{\max}$ so that it has the series expansion 
    \begin{equation*}
        w(\alpha,\beta,\gamma) = \sum_{n=0}^{n_\text{max}}\sum_{m=-n}^n \sum_{\mu=-n}^n\acoef{n}{m\mu} \WD{n}{\mu}{m}(\alpha,\beta,\gamma).
    \end{equation*} 
    Let $w \in \mbb{C}^M$ be the vector of $M$ measurements of the periodic extension of $w(\alpha, \beta, \gamma)$ taken on even Nyquist grid in $\tthre$ ($2n_{\max}+2$ samples in each dimension). Assume these samples are corrupted by the additive measurement noise, $\eta\in \mbb{C}^M$. Then the coefficients of the Wigner $D$-function series expansion, $\acoef{n}{m\mu}$, can be estimated using the following two step method:
    
    \begin{enumerate}
        \item Solve for the vector $b\in \mbb{C}^{N_F}$ with $N_F = (2n_{\max}+2)^3$ using the linear inverse problem
        \begin{align}
            N_F^{-1/2} w &= U_F b + N_F^{-1/2} \eta \label{eq:3ddft_problem},
        \end{align}
        where $U_F\in\mbb{C}^{N_F \times N_F}$ is the unitary matrix representing the normalized 3DDFT having $2n_{\max}+2$ samples in each dimension.
        \item Using $b$ from step 1, solve for the coefficients $\acoef{n}{m\mu}$ by solving the linear inverse problem
        \begin{equation} \label{eq:b_inv_prob}
            b = B a,
        \end{equation}
        where $a$ is the vector of $\acoef{n}{m\mu}$ and $B$ can be written as a block matrix with blocks $B^{m\mu}$ that only operate on the coefficients $\acoef{n}{m\mu}$ with fixed $m$ and $\mu$.
    \end{enumerate}
\end{lemma}

\textit{Proof:} See \hyperref[app:ProofOfMainLem]{Appendix}.

With Lemma~\ref{lem:dft_cs_for_wD_series} setup so that the inverse problem in \eqref{eq:wigner_D_expanison} is reformulated as a linear inverse problem with a unitary measurement matrix, \eqref{eq:3ddft_problem}, and the auxiliary problem \eqref{eq:b_inv_prob},  we state the compressive sampling guarantee for the problem the Fourier basis, \eqref{eq:3ddft_problem}. The SW/SH version of this result is stated in \Cref{rem:sh_cs}.

\begin{theorem}[Sparse Recovery for Spherical Field Measurements Using a Sub-sampled 3DDFT] \label{thm:subsampled_3ddft}
    Consider the linear inverse problem specified in \eqref{eq:3ddft_problem}. Suppose that $\|N_F^{-1/2} \eta\|_\infty \leq \epsilon$, $P_\Omega$ is the matrix that selects a subset $\Omega$ of the rows of $U_F$. If $\Omega$ is selected uniformly at random from all subsets of size $M$ with
    \begin{equation}
        M\geq C_2 s_F\ln^4(N_F),
    \end{equation}
    then with probability $1-(N_F)^{-\ln^3(N_F)}$, if $\widehat{b}'$ is the solution to 
    \begin{equation}
        \widehat{b}' = \arg \min_{z\in\mathbb{C}^{N_F}}  \|z\|_1 \; \text{subject to} \; \|N_F^{-1/2} w-P_\Omega U_F z\|_2 \leq \sqrt{M}\epsilon
    \end{equation}
    then
    \begin{equation}\label{$T^3$-SF_cs_err}
        \left\|b'-\widehat{b}'\right\|_2 \leq C_1\left( \frac{\sigma_{s_F}(b')_1}{\sqrt{s_F}} + \epsilon \right).
    \end{equation}
    Here $C_1\geq0$ and $C_1$ only depends on the restricted isometry constant of $P_\Omega U_F$, $\delta_{2s_F}$.
\end{theorem}

\textit{Proof:} Note that $U_F$ can be written as a Kronecker product of one-dimensional DFT matrices, $U_{DFT}\in \mbb{C}^{2n_{\max}+2}$. That is, $U_F = U_{DFT} \otimes U_{DFT} \otimes U_{DFT}$. The elements of $U_{DFT}$ are of the form $e^{-\mi \theta }/\sqrt{2n_{\max}+2}$ with $\theta\in\mbb{R}$. Thus, the elements of $U_F$ satisfy $\left|[U_F]_{ij}\right| = \left| e^{-\mi (\theta_1+\theta_2+\theta_3)} / \sqrt{(2n_{\max}+2)^3} \right| \leq N_F^{-1/2}$. Pairing this fact with the unitarity of $U_F$ and then using Corollary \ref{cor:unitary_rip} and Theorem \ref{thm:sparse_recovery_rip} gives the desired result. \bsquare

\begin{remark} \label{rem:wigner_d_sparsity_vs_fourier}
    Changing our problem in~\eqref{eq:wigner_D_expanison} from the Wigner $D$-function basis to the Fourier basis results in a change in sparsity. This is because multiplying by $\t{B}{^{m\mu}}$ sums the entries of $\acoef{n}{m\mu}$ along $n$. Typically, due to symmetries of the field and probe, a device's coefficients only use a few $m,\;\mu$ subspaces. If the number of such subspaces used is $n_{m\mu}$, then the worst-case sparsity of $b'$, $s_F$, is $(2n_{\max}+2)n_{m\mu} = ({N_F})^{1/3}n_{m\mu}$. So the required number of measurements is 
    \begin{equation}
        M\geq C_2 N_F^{1/3} n_{m\mu} \ln^4(N_F).
    \end{equation}
    Moreover, we know $n_{m\mu}\leq s_D$, where $s_D$ is the sparsity in the Wigner $D$-function basis. Noting that we can relate $N_F$ to $N_D$ as $N_F \leq C' N_D$ (set $C'=6$ for example), we can also use the condition 
    \begin{equation} \label{eq:fourier_m_in_d_sparsity}
        M\geq C_2 (C'N_D)^{1/3} s_D \ln^4(C'N_D).
    \end{equation}
    When comparing the scaling of $M$ in the Fourier basis to the Wigner $D$-function basis \eqref{eq:wigner_d_m_scaling}, we have (ignoring log factors) $N_D^{1/3}$ and $N_D^{1/6}$, respectively. By going to the Fourier basis we gain a factor of $N_D^{1/6}$, which is slightly worse. However, the method presented here does not require sampling from arbitrary points on $\sot$; it sub-samples the Nyquist grid, which is much easier for measurement devices to achieve. As a further note, the structure of the $\t{B}{^{m\mu}}$ is such that this worst case increase in sparsity when transforming from the Wigner $D$-function to Fourier basis is attained only when $\acoef{n}{m\mu}\neq 0$ for $n=n_{\max}$. This is typically not the case, as larger $\acoef{n}{m\mu}$ tend to be at lower $n$, so the bound $n_{m\mu}\leq s_D$ is likely loose. Thus, in practice, we might expect better a better scaling of $M$ than what we see in \eqref{eq:fourier_m_in_d_sparsity}.
\end{remark}

\begin{remark}
    For emphasis, we compare the result of Theorem \ref{thm:subsampled_3ddft} to the classical Nyquist sampling approach in EM, which uses a $\mu=\pm1$ probe. In the classical approach, the number of measurements must scale with $N_D^{2/3}$. The result in~\eqref{eq:fourier_m_in_d_sparsity} requires that $M$ scales with $N_D^{1/3}$ times log factors. Ignoring the constants and log factors, this beats the classical $\mu=\pm1$ Nyquist approach by a factor of $N_D^{1/3}$. Importantly, the sampling required here is to take a size $M$ subset of the grid on $\tthre$ (and so $\sot$). This requires accessing a subset of the positions the classical approach uses on the sphere enclosing a device, unlike the results in~\cite{bangunSensingMatrixDesign2020}, which require arbitrary positions.
\end{remark}

\begin{remark}\label[remark]{rem:fourier_grid_size}
    In the proof of Lemma~\ref{lem:dft_cs_for_wD_series}, we extended the limits of the Fourier series indices to contain an even number of frequencies in $m$, $\mu$, and $m'$. The purpose of this is so that the Nyquist sampling grid given by $(\alpha_j, \beta_k, \gamma_l) = (2\pi j/(2n_{\max}+2),\;2\pi k/(2n_{\max}+2),\;2\pi l/(2n_{\max}+2))$ with $j,k,l\in\{-n_{\max}-1,-n_{\max},\cdots,n_{\max}\}$ results in each measurement on $\sot$ corresponding to two or more measurements on $\tthre$. This implies that satisfying the bound on $M$ in \Cref{thm:subsampled_3ddft} requires the number measurements on $\sot$ to be at most $M/2$. Contrast this with an odd grid, which has no repeated points and thus the number of measurements needed on $\sot$ is $M$. In more detail, the even grid results in the points on the poles $\beta=0$ or $\beta = \pi$ being sampled, thus there is a degeneracy in the choice of non-polar angles $\alpha_j$ and $\gamma_l$ at the poles. At $\beta=0$ any points with $\alpha_j + \gamma_l = \rho$ for fixed $\rho$ represent the same physical measurement. The condition for $\beta=\pi$ is $\alpha_j + \pi - \gamma_l = \rho$ for a fixed $\rho$. This fact results in $2n_{\max}+2$ repeated points in $\tthre$ when a pole is measured in $\sot$. Any other non-polar points have two repeated measurements where $(\alpha_j, \beta_k, \gamma_k)$ is the same measurement as $(\alpha_j+\pi, -\beta_k, \gamma_k-\pi)$. 
\end{remark}

\begin{remark}\label[remark]{rem:sh_cs}
    Consider Lemma \ref{lem:dft_cs_for_wD_series} and take the special case where one has a set of SW/SH coefficients. Assuming that the field is measured ideally at a point amounts to setting all $\acoef{m'}{m\mu}$ with $\mu\neq 0$ or $m\neq 0$ to zero. In this case, (ignoring the noise) the measurements $w(\alpha_j, \beta_j,\gamma_j)$ depend only on $(\beta_k,\gamma_l)$ or $(\alpha_j,\gamma_k)$, respectively. The result is that we can consider this special case as a two-dimensional Fourier series after appropriate normalization. Thus, using nearly the same analysis that results in \Cref{thm:subsampled_3ddft}, we arrive at being able to use QCBP to solve for the non-zero entries in $b$ from a sub-sampled 2DDFT so long as 
    \begin{equation}\label{eq:M_2DDFT_CS}
        M \geq C_2 s_F \ln^4(N_{F,2D}),
    \end{equation}
    where $N_{F,2D} = (2n_{\max}+2)^2$ is the number of basis functions in the band-limited 2DDFT. Carrying out the same analysis as the above remarks and noting the worst case $s_F$ is $2n_{\max}+2 = \sqrt{N_{F,2D}} \leq \sqrt{C''N_{SH}}$ (using say $C''=2$) times the number of $m$ ($\mu$) subspaces used, which is at most $s_{SH}$, we can also use
    \begin{equation}\label{eq:M_2DDFT_CS}
        M \geq C_2 \sqrt{C''N_{SH}} s_{SH} \ln^4(C''N_{SH}).
    \end{equation}
    As with the full $\sot$ case, using the Fourier basis Nyquist grid on the sphere for CS results in an increase in the number of measurements required when compared to the best case scaling for SW/SH series of $O(N_{SH}^{1/6})$. Again, however, the Fourier method does not require being able to sample the sphere at any arbitrary point like the method achieving $O(N_{SH}^{1/6})$ does.
\end{remark}

\section{Examples and Numerical Investigations}\label{sec:numerical_examples}
\subsection{Analysis of Basis Transformation} \label{sec:rand_sparsity}
The transformation from the Wigner $D$-function to Fourier basis in Lemma \ref{lem:dft_cs_for_wD_series} will affect the sparsity of the coefficients. From the form of the transformation we can see that this mapping is likely to increase the sparsity, but only within already populated $m,\;\mu$ subspaces. This becomes even more apparent if we look at the structure of the entries of $\t{B}{^{m\mu}}$. 

\begin{figure}[ht]
    \centering
    \includegraphics[width=\reprintcolumnwidth]{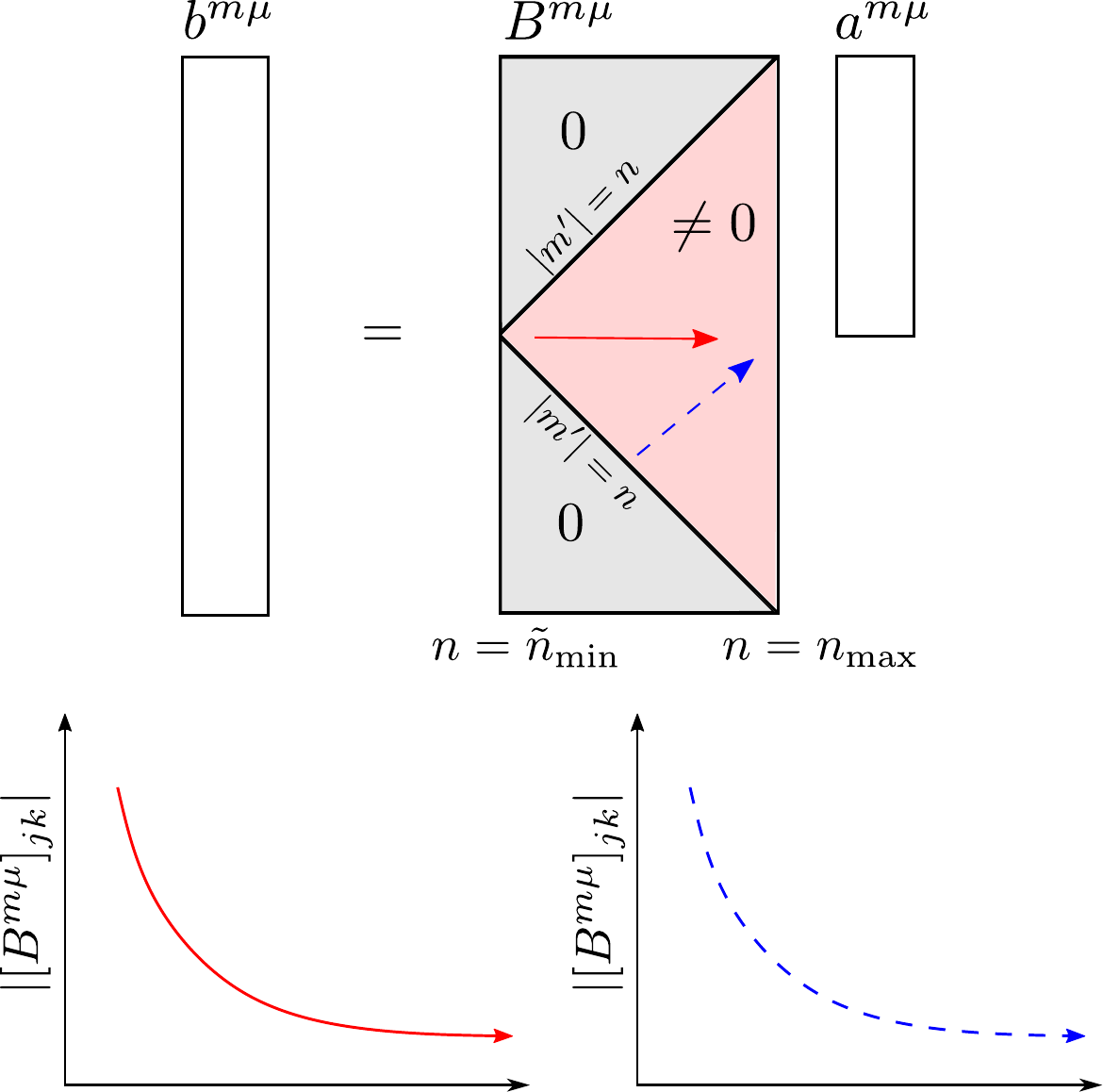}
    \caption{\tit{$\t{B}{^{m\mu}}$ Structure.} Depiction of the non-zero entries of the transformation matrix from the Wigner $D$-function basis to the Fourier. The triangle-like structure helps visualize how sparsity can change between bases.} \label{fig:B_matrix_structure}
\end{figure}

Investigating many examples of $\t{B}{^{m\mu}}$ for a different $m,\;\mu$, and $n_{\max}$ reveals a common structure these matrices share. This structure can be seen in \Cref{fig:B_matrix_structure}, which shows that this structure is triangle-like, with zero entries in the grey regions and non-zero entries in the region containing the arrows. This structure is manifest in the proof of Lemma \ref{lem:dft_cs_for_wD_series} (see \eqref{eq:B_m_mu_entires} in the \hyperref[app:ProofOfMainLem]{Appendix}). However, this equation does not give straightforward insight into how the non-zero elements behave. Our investigations show that the overall trend is given by the two curves in the bottom of \Cref{fig:B_matrix_structure} paired with their corresponding direction in the depiction of the matrix. Pairing these diagrams we see that rows tend to decrease as we increase along the columns (see the horizontal line) but not to zero. We also see that elements along the lines parallel to the $|m'|=n$ line also decay, but not to zero.

The triangle-like structure of the $\t{B}{^{m\mu}}$ also implies that we can make statements about the sparsity $s_F$ in the Fourier basis if we know properties of the $\acoef{n}{m\mu}$ in each $m,\;\mu$ subspace, i.e., properties about the symmetry of the field or the spatial variation of it. For example, if we know that the different $m,\;\mu$ subspaces have different cutoffs in $n$, which we can label $n_{\max}^{m\mu}$, then the sparsity in the Fourier basis satisfies the bound
\begin{equation}
    s_F \leq \sum_{m,\mu:\; \acoef{n}{m\mu}\neq 0} (2n_{\max}^{m\mu} + 1).
\end{equation}
As a note, the different $n_{\max}^{m\mu}$ for each $m,\;\mu$ subspace can be seen as smoothness criteria on the function $w(\alpha,\beta,\gamma)$ if one takes the $n_{\max}^{m\mu}$ to be decreasing with increasing $|m|$ and $|\mu|$.

\subsection{Analysis of Sparsity Change with Random Coefficients}
\begin{figure}[ht]
    \centering
    \includegraphics[width=\reprintcolumnwidth]{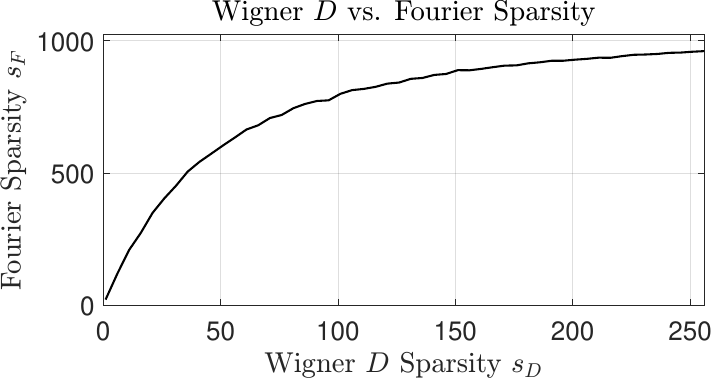}
    \caption{\tit{Average $s_F$ Versus $s_D$.} The average Fourier sparsity level, $s_F$, as a function of Winger $D$-function basis sparsity level, $s_D$. Wigner $D$-function coefficients are randomly set to one and transformed to the Fourier basis where the sparsity level is averaged over 100 trials.} \label{fig:average_sparsity_transformation}
\end{figure}

To get more intuition behind how the Fourier sparsity $s_F$ depends on the Wigner $D$-function sparsity $s_D$ in general, we numerically test this dependence as the basis is changed with a band-limit of $n_{\max}=15$ for ease of computation. In particular, we look at an analog of the case where an ideal probe is used. This choice of probe means that the Wigner $D$-function coefficients are only non-zero for $\mu = 0$. For our analogous setup, we pick a sparsity $s_D$, uniformly at random set $s_D$ coefficients $\acoef{n}{m0}=1$, and then transform to the Fourier basis and calculate $s_F$. This is repeated for 100 trials with each value of $s_D$ and averaged over all trials for a given $s_D$. The results are plotted in \Cref{fig:average_sparsity_transformation}. While this is not exactly representative of typical spherical field measurements, since the $\acoef{n}{m\mu}$ typically will not all be equal, \Cref{fig:average_sparsity_transformation} shows how $s_F$ can begin to saturate (the maximum possible $s_F$ is $1024$) and become problematic for CS with lower $s_D$ when there is no structure to the coefficients. As a note, this experiment is likely to fare worse than real devices; most devices of interest have some amount of symmetry or structure in regards to which $m, \mu$ subspaces are nonzero.

\subsection{Ideal and Non-Ideal Probe Measurements of Directed Speaker} \label{sec:speaker_sparsity}
In actual measurement systems, it is typical to use devices that are as ideal as possible. In many acoustic systems this means an ideal probe is used and the field is measured directly. Such probes are only sensitive to $\mu=0$ SW modes in its coordinate system. In the EM case, using an ideal probe means that the probe is only sensitive to $\mu=\pm 1$ SWs. Moreover, among the fields that one characterizes using spherical field measurements, it is common to see fields with varying levels of symmetry. For example, the sound field radiated from a box speaker, transducer, or a spherical array of transducers may have varying levels of rotational symmetry along their main beam. If this main beam is aligned along the azimuthal axis of the measurement system, then the more symmetric the field is, the fewer number of $m\neq 0$ SW modes will be needed. With this idea in mind, we investigate the change in sparsity transforming from the Wigner $D$-function to the Fourier basis for the coefficients of an example speaker at three frequencies:
\begin{enumerate}
    \item \tbf{Case 1 (C1)}: 1098 Hz
    \item \tbf{Case 2 (C2)}: 1400 Hz
    \item \tbf{Case 3 (C3)}: 1895 Hz
\end{enumerate}
These frequencies are selected so that they provide various levels of axial symmetry as determined by the fractional contribution the speaker's $m\neq 0$ SW modes make to the square of the $\ell_2$ norm of the SW coefficients.

\begin{figure}[ht]
    \centering
    \figcolumn{
        \fig{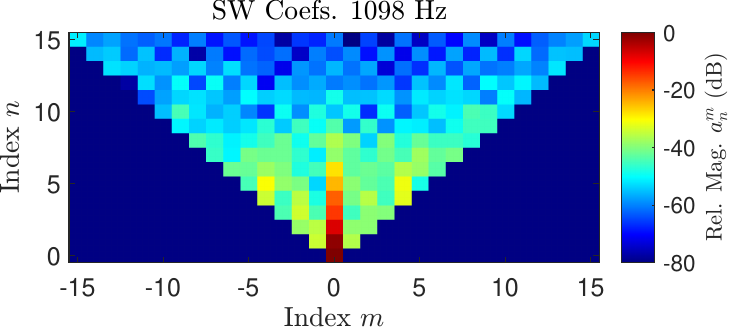}{\reprintcolumnwidth}{(a)}
        \fig{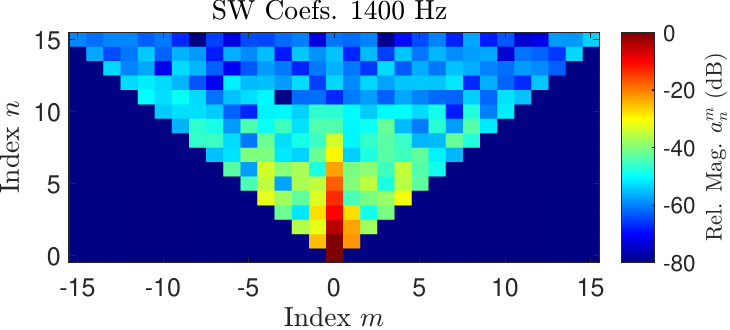}{\reprintcolumnwidth}{(b)}
        \fig{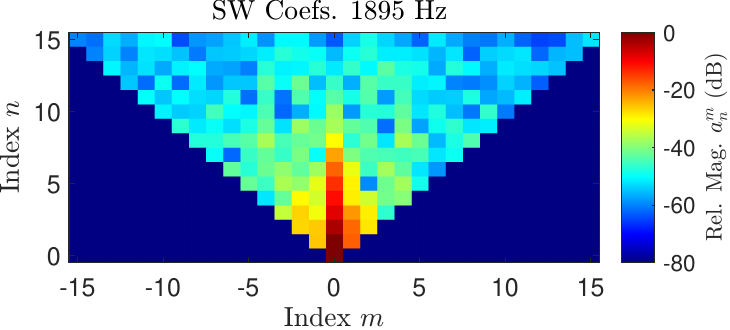}{\reprintcolumnwidth}{(c)}
    }
    \caption{{\it Source Spherical Wave Coefficients.} Spherical wave coefficients magnitudes up to the band-limit $n_{\max}=15$ for the IEM loudspeaker cube driver 1 at 1098 Hz (a), 1400 Hz (b), and 1895 Hz (c).}
    \label{fig:sw_coefs}
\end{figure}

The specific speaker we consider is driver 1 of the IEM Loudspeaker Cube~\cite{meyer-kahlenDesignMeasurementFirstOrder2018}, for which directivity measurement data is openly available at~\cite{iemDirPatDatabaseVisualization2017}. SH coefficients for the speaker at various frequencies are calculated using the open-source code made available by Ahrens et. al.\ at~\cite{ahrensDatabaseSphericalHarmonic2020, ahrensComputationSphericalHarmonics2020} (which also contains the directivity measurement data for the IEM Loudspeaker Cube). This code fits the loudspeaker measurements to a SH series with band-limit $n_{\max} = 17$. For ease of simulation we truncate this data to $n_{\max}=15$. From the SH coefficients, we calculate the SW coefficients $\acoef{n}{m}$ by dividing out the appropriate spherical Hankel function evaluated at the distance between the probe coordinate system and the speaker coordinate system, $r_{ab}=0.75$m. In the original data output from the code in ~\cite{ahrensDatabaseSphericalHarmonic2020}, the main beam of the speaker is along the negative $y$ axis. To get the beam along the azimuthal ($z$) axis of the spherical measurements, we rotate the output coefficients using Wigner $D$-functions~\cite{varshalovichQuantumTheoryAngular1988}. As a note, we normalize the coefficients to have an $\ell_2$ norm of $1$, i.e., $\sum_{n,m} |\acoef{n}{m}|^2=1$. The relative magnitudes of the SW coefficients in dB,
\begin{equation}
    \textnormal{Rel. Mag.}\; \acoef{n}{m} = 20\log_{10}\left(\frac{|\acoef{n}{m}|}{\max_{n,m} |\acoef{n}{m}|}\right),
\end{equation}
are presented in \Cref{fig:sw_coefs}. As can be seen in \Cref{fig:sw_coefs}, the contribution of the $m\neq 0$ modes to the $\ell_2$ norm squared increases with frequency. In particular, \tbf{C1} has 0.45\% of the signal's $\ell_2$ norm squared in the $m\neq 0$ modes, \tbf{C2} has 1.05\%, and \tbf{C3} has a contribution of 2.22\%.

The Wigner $D$-function coefficients for the speaker, $\acoef{n}{m\mu}$, are a product of the SW coefficients of the source, $\acoef{n}{m}$, and the response constants, $\Ccoef{n}{\mu}$. That is, $\acoef{n}{m\mu} = \acoef{n}{m}\Ccoef{n}{\mu}$. Since the response constants also have an effect on the sparsity and are a result of the probe, we will test the change in sparsity from the Wigner $D$-function basis to the Fourier basis with an axisymmetric probe as well as non-axisymmetric probes. The baseline probe we select is the ideal probe. In the axisymmetric case for each frequency, the response constants are taken to be ideal for the 1098 Hz signal and are equal to $\Ccoef{n}{0}=\frac{\sqrt{2n+1}}{4\pi} h_n^{(1)}(k r_{ab})$ with the remainder being zero. Here $h_n^{(1)}$ is the Spherical Hankel function of the first kind, and $k$ is the wavenumber of the 1098 Hz sound field. We use this same probe for all fields so we avoid introducing variations in the $\ell_2$ norm of the Wigner $D$-function coefficients by changing the probe. For non-ideal probe measurements, we assume the non-ideal nature of the probe comes from the probe being more sensitive to high-order $\mu$ modes at two increased levels (specified below). This sensitivity is set to be randomly selected. Thus, \tbf{C1}-\tbf{C3} will each have three sub-cases where the response constants will be:
\begin{enumerate}[label=(\alph*)]
    \item $\Ccoef{n}{0}=\frac{\sqrt{2n+1}}{4\pi} h_n^{(1)}(k r_{ab})$, $\Ccoef{n}{m}=0$ otherwise.
    \item $\Ccoef{n}{0}=\frac{\sqrt{2n+1}}{4\pi} h_n^{(1)}(k r_{ab})$, $\Re(\Ccoef{n}{\pm1}), \Im(\Ccoef{n}{\pm1}) \sim \mathcal{N}(0, 0.01\max_{n}|\Ccoef{n}{0}|)$, $\Ccoef{n}{m}=0$ otherwise.
    \item $\Ccoef{n}{0}=\frac{\sqrt{2n+1}}{4\pi} h_n^{(1)}(k r_{ab})$, $\Re(\Ccoef{n}{\pm1}), \Im(\Ccoef{n}{\pm1}) \sim \mathcal{N}(0, 0.01\max_{n}|\Ccoef{n}{0}|)$,  $\Re(\Ccoef{n}{\pm2}), \Im(\Ccoef{n}{\pm2}) \sim \mathcal{N}(0, 0.001\max_{n}|\Ccoef{n}{0}|)$, $\Ccoef{n}{m}=0$ otherwise.
\end{enumerate}

To see the changes in sparsity for all of the above cases, we compare the non-zero coefficients (to floating point precision) sorted largest to smallest in both the Wigner $D$-function basis as well as the Fourier basis, where each set is normalized with respect to the largest coefficient in the given basis. Explicitly, we plot the $\mrm{Coefficient\;Relative\;Magnitude}$ in dB given by 
\begin{equation}
    \textnormal{Coefficient Relative Magnitude} = 20\log_{10}\left(\frac{|c_j|}{|c_1|}\right),
\end{equation}
where the $c_j,\;j=1,2,\cdots$ are the sorted coefficients in either the Wigner $D$-function basis or the Fourier basis. Along with the coefficients, it is also informative to investigate the effect of keeping only the $n_c$ largest coefficients in a given basis. To that end, we also plot the coefficient error normalized by the actual coefficient $\ell_2$ norm squared in dB,
\begin{equation}
    \textnormal{Normalized Error}(c,n_c) = 10 \log_{10} \left(\frac{\|c_{n_c}-c\|^2}{\|c\|^2}\right),
\end{equation}
where $c$ is a vector of coefficients and $c_{n_c}$ is the vector of with all but the $n_c$ largest coefficients set to zero. We plot the normalized error in the Wigner $D$-function basis, in the Fourier basis, and then in the Wigner $D$-function basis after keeping $n_c$ coefficients in the Fourier basis and transforming back to the Wigner $D$-function basis.  

\begin{figure}[ht]
    \centering
    \figcolumn{
        \fig{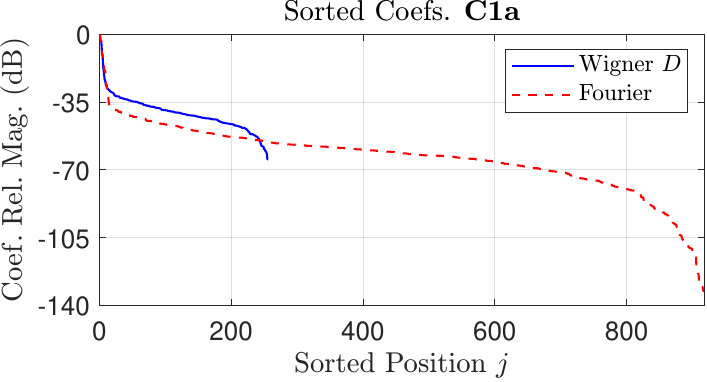}{\reprintcolumnwidth}{(a)}
        \fig{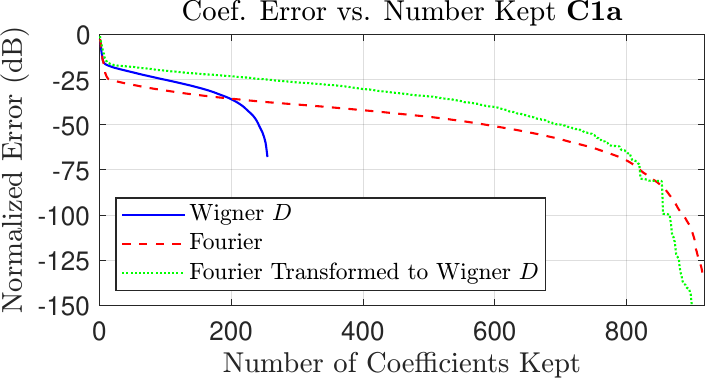}{\reprintcolumnwidth}{(b)}
    }
    \caption{{\it Sorted Coefficients and Normalized Error Case \tbf{C1a}.} The coefficient relative magnitude, (a), in the Wigner $D$-function and Fourier bases rapidly decays initially, indicating compressibility in either basis. The normalized error, (b), drops below $-30$ dB before all coefficients are kept in each case, indicating CS recovery with smaller sparsities $s_F$ should give accurate results.} \label{fig:sparsities_c1a}
\end{figure}

\begin{figure}[ht]
    \centering
    \figcolumn{
        \fig{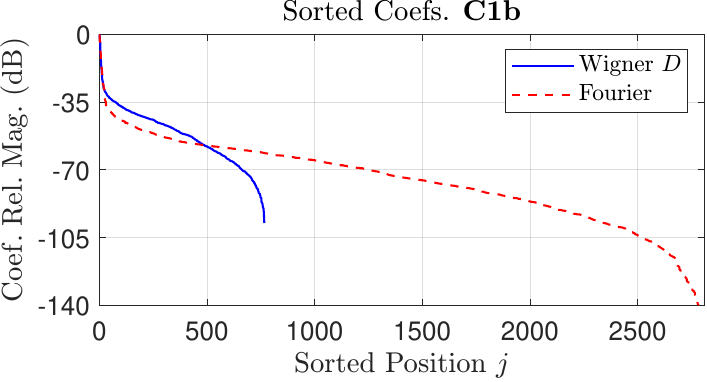}{\reprintcolumnwidth}{(a)}
        \fig{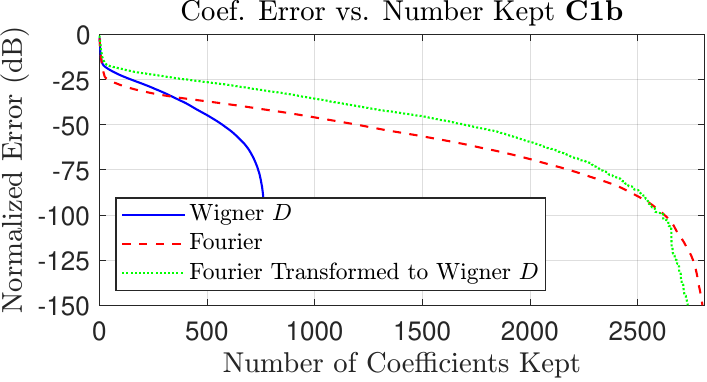}{\reprintcolumnwidth}{(b)}
    }
    \caption{{\it Sorted Coefficients and Normalized Error for Case \tbf{C1b}.} The coefficient relative magnitude, (a), and the normalized error, (b), show similar trends to those seen in~\Cref{fig:sparsities_c1a}, so the same conclusions apply. Any scaling in the number of nonzero coefficients and error decay proportional to the increased asymmetry of the selected probe.} \label{fig:sparsities_c1b}
\end{figure}

\begin{figure}[ht]
    \centering
    \figcolumn{
        \fig{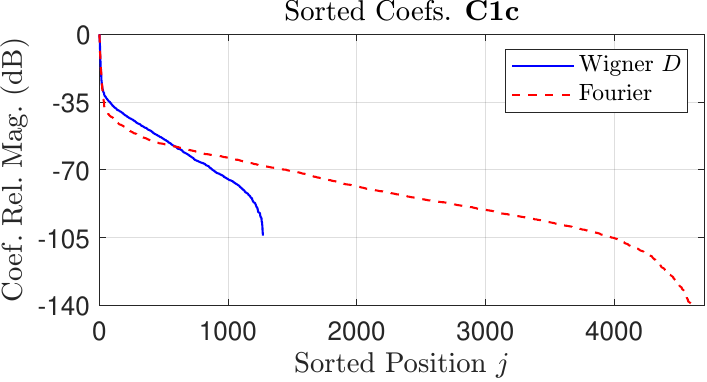}{\reprintcolumnwidth}{(a)}
        \fig{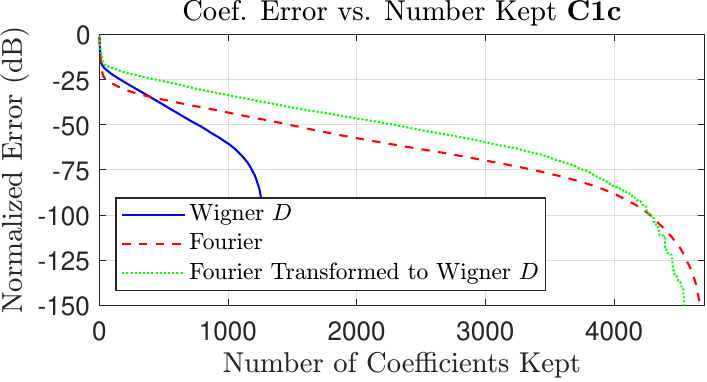}{\reprintcolumnwidth}{(b)}
    }
    \caption{{\it Sorted Coefficients and Normalized Error for Case \tbf{C1c}.} Again, the coefficient relative magnitude, (a), and the normalized error, (b), show trends similar to those seen in~\Cref{fig:sparsities_c1a,fig:sparsities_c1b}. Thus, the same conclusions apply. Scaling in the number of nonzero coefficients and error decay is again proportional to the increased asymmetry of the selected probe.} \label{fig:sparsities_c1c}
\end{figure}

The sorted coefficients and normalized errors for cases \tbf{C1a-c} can be seen in \Cref{fig:sparsities_c1a,fig:sparsities_c1b,fig:sparsities_c1c}. For the sake of brevity in the main text, the corresponding plots for cases \tbf{C2a-c} and \tbf{C3a-c} are included as supplementary files\footnote[1]{See supplementary
material at [URL will be inserted by AIP] for plots of the sorted coefficients and normalized errors in \tbf{C2a-c} and \tbf{C3a-c} (defined in \Cref{sec:speaker_sparsity})}. If we compare the number of non-zero coefficients for a fixed frequency but increasing probe asymmetry, e.g., \tbf{C1a-c} in \Cref{fig:sparsities_c1a,fig:sparsities_c1b,fig:sparsities_c1c}, then we see the number of non-zero coefficients increases by a factor of $\approx 3$ from case \tbf{a} to \tbf{b} and then a factor of $\approx 5$ from \tbf{a} to \tbf{c}, regardless of basis. These scaling factors are in line with the scaling in the number of non-zero response constants $\Ccoef{n}{m}$ between the cases. Thus, ideally the asymmetry of any probe used is small so that the additional non-zero coefficients induced will be smaller than the SW coefficients one is trying to recover. This trend continues for \tbf{C2a-c} and \tbf{C3a-c} in the supplementary material\footnote[1]~. Comparing the number of non-zero coefficients as we increase sound frequency but keep the same response constants by using \Cref{fig:sparsities_c1a,fig:sparsities_c1b,fig:sparsities_c1c} and the supplementary material\footnote[1]~, we observe that the number of non-zero coefficients in the Fourier basis is approximately three times that of the Wigner $D$-function basis for each case. These results are much better than the worst case, which would be an increase by a factor of $\approx 2n_{\max}+1$ (since one coefficient per $m,\mu$ subspace in the Wigner $D$-function basis can map to $2n_{\max}+1$ in the Fourier basis).

In terms of the normalized error, in \Cref{fig:sparsities_c1a,fig:sparsities_c1b,fig:sparsities_c1c} and the supplementary material\footnote[1]~ we see that the errors drop below $-30$ dB well before all coefficients are being kept. The most important curve is the green dotted curve (keeping $n_c$ coefficients in the Fourier basis and transforming back to the Wigner $D$-function basis), since this is the most relevant number for the method we propose. This implies that the coefficients in the Fourier basis are compressible and CS recovery with smaller sparsities $s_F$ should be reasonably accurate. In fact, in all cases we see a very rapid drop to near $-15$ dB or better in the first $n_c \approx20$ coefficients, with slower gains in accuracy after that. Lastly, we note that the Fourier transformed to Wigner $D$-function curves (green dotted) have a larger tail. This is as expected from the broadened tail we see when looking at the sorted coefficients in the Fourier bases. 

\subsection{Compressive Sampling Recovery}\label{sec:cs_recovery_numerics}
In this section, we demonstrate the recovery of the sound field emitted by the IEM Loudspeaker Cube driver 1 at 1098 Hz by solving the two-step problem in~\eqref{eq:fourier_problem} and~\eqref{eq:a_b_coef_problem} using compressive sampling according to Theorem~\ref{thm:subsampled_3ddft}. For this demonstration, we simulate noiseless measurements of the sound field emitted the IEM Loudspeaker Cube driver 1 at 1098 Hz taken by an ideal probe on the sphere of radius $r_{ab} = 0.75$ m. Using the ideal probe implies that measurements $w$ collapse to the field value at a position on the sphere specified by $(\alpha, \beta, \gamma)$ according to \Cref{fig:rotation_diagram} and the associated discussion in~\Cref{sec:background} (see \Cref{sec:background} also for a brief discussion of this collapsing). Thus, measurements are simulated by straightforwardly calculating the field value for a given position on the sphere by using the SW coefficients extracted as described in~\Cref{sec:speaker_sparsity}. As a further note, using an ideal probe implies that the coefficients $\acoef{n}{m\mu}$ are zero for all $\mu$ except $\mu=0$. Since these coefficients are zero and the DFT associated with the $\mu$ index is no longer needed, the 3DDFT from Theorem \ref{thm:subsampled_3ddft} is reduced to a 2DDFT as discussed in \Cref{rem:sh_cs}. 

\begin{figure}[ht]
    \centering
    \figcolumn{
        \fig{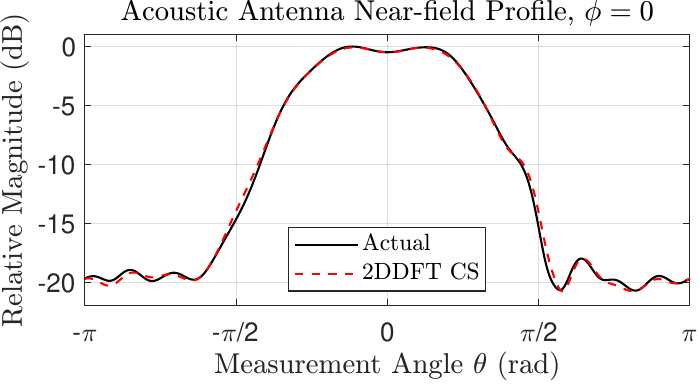}{\reprintcolumnwidth}{(a)}
        \fig{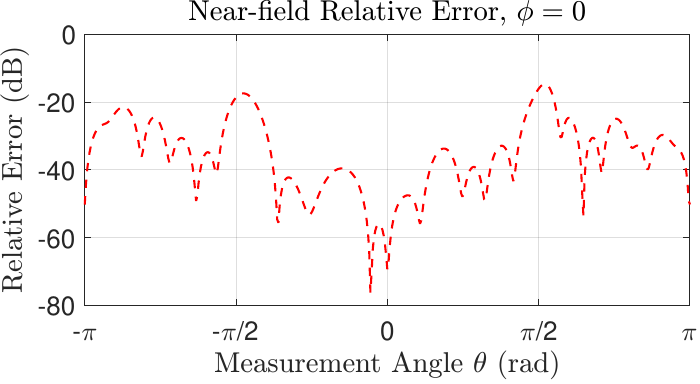}{\reprintcolumnwidth}{(b)}
    }
    \caption{\tit{Near-field Reconstruction}. Example field reconstruction for case \tbf{C1a}, (a), and relative error, (b), using the sub-sampled 2DDFT. This reconstruction uses a 306 unique measurements while classical Nyquist reconstruction would require 496 measurements.}\label{fig:nf}
\end{figure}

We can compare the above reconstruction approach with classical Nyquist-based measurements and the algorithm from acoustics~\cite{wittmannProbecorrectedSphericalNearfield1992}. In this classical algorithm, the number of measurements is dictated by the Nyquist sampling theorem and, for the band-limit $n_{\max}=15$, requires a minimum of 496 measurements for a perfect reconstruction with no measurement noise~\cite{wittmannProbecorrectedSphericalNearfield1992}. In this experiment we randomly select 400 measurement positions, which results in 306 unique physical measurements (due to random selections repeating points on $\sot$, see \Cref{rem:fourier_grid_size}). When measurements are noiseless, Basis Pursuit, not QCBP, is used. The reconstructed sound field along with the original and the relative error along the $\phi=0$ axis are shown in \Cref{fig:nf}. The reconstructions are plotted in terms of magnitude relative to the maximum actual field in dB, 
\begin{equation}
    \text{Relative Magnitude} = 20\log_{10}\left(\frac{\left|\widehat{F}\right|}{\max_{\alpha,\beta,\gamma}|F|}\right),
\end{equation} and the Relative Error is given in dB as
\begin{equation}\label{eq:coef_rel_err}
    \text{Relative Error} = 20\log_{10}\left(\frac{\left|F-\widehat{F}\right|}{|F|}\right).
\end{equation}
As can be seen, the sub-sampled 2DDFT performs well with the relative error near $-30$ dB over most of the $\theta$ range. As a note, using CS with SHs and random measurements on the sphere can nearly halve the number of measurements reduced and maintain similar accuracy, as expected when comparing \Cref{thm:subsampled_3ddft} to~\cite{rauhutSparseLegendreExpansions2012}. However, the sub-sampled 2DDFT method has the advantage of being sampled from a sub-selection of the Nyquist grid on the sphere.

\subsection{Compressive Sampling Recovery Versus Measurement Number}
\begin{figure}[ht]
    \centering
    \includegraphics[width=\reprintcolumnwidth]{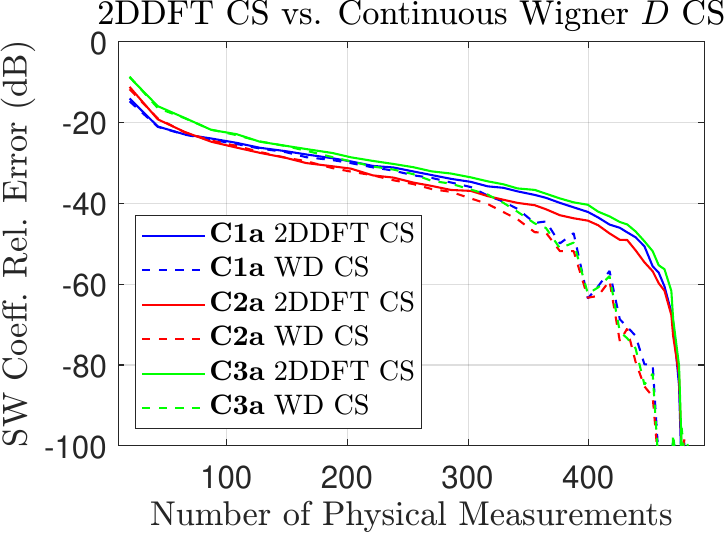}
    \caption{{\it Relative Error Versus Measurement Number in Cases 1a, 2a, and 3a for 2DDFT CS and Continuous Wigner $D$-function CS.} For each number of measurements, the relative error is averaged over 25 trials. Both compressive sampling methods perform similarly for cases \tbf{C1a}, \tbf{C2a}, and \tbf{C3a} until about 300 measurements, where the Wigner $D$-function method begins to perform better. By this sample number, the Wigner $D$-function method is close to fully determined while the Fourier method is still under-sampled.}\label{fig:recon_error_versus_meas}
\end{figure}

In this section, we investigate CS recovery using the 2DDFT method developed in this paper compared to the continuous Wigner $D$-function approach developed in~\cite{bangunSensingMatrixDesign2020} with $\mu = 0$ functions only (i.e., re-scaled spherical harmonics). We do this by using each method to solve for the coefficients in cases \tbf{C1a}, \tbf{C2a}, and \tbf{C3a} as we vary the number of measurements used. In particular, we look at the relative error in the recovered partial wave coefficients given by
\begin{equation}
    \textnormal{Relative Error} = \frac{\sum_{n,m}^{n_{\max}} |a_n^m-\widehat{a}_n^m|^2}{\sum_{n,m}^{n_{\max}} |a_n^m|^2}.
\end{equation}
Here, the $\ahcoef{n}{m}$ are the recovered SW coefficients and $a_n^m$ are the actual SW coefficients. Again, as the number of randomly selected measurements and the number of unique ``physical measurements'' (simulated measurements corresponding to unique positions on the sphere) will be different, we plot the Relative Error as a function of the average number of unique physical measurements over 25 trials while varying the number of randomly selected measurements in the Fourier domain. The results of this are in \Cref{fig:recon_error_versus_meas}. For each case, the 2DDFT CS and the continuous Wigner $D$-function CS approaches perform similarly when the number of physical measurements is less than 300. Over 300 measurements, however, the Wigner $D$-function method improves relative to the 2DDFT CS method. This can be explained by the fact that the Wigner $D$-function method has a smaller dimension for the coefficient space, so with high sample numbers the problem can be close to fully determined. Note, as the number of ``physical measurements'' approaches the fully sampled Nyquist grid (496 measurements), the accuracy of the reconstruction becomes nearly perfect for the 2DDFT CS approach, as expected.

The approach in Lemma \ref{lem:dft_cs_for_wD_series} and the statement of \Cref{thm:subsampled_3ddft} do not explicitly allow for sampling grids that are denser than the Nyquist rate as they are written. However, a straightforward generalization of \Cref{cor:unitary_rip} allows for a measurement matrix that is a selection of columns from a unitary matrix in a larger dimension. For example, one can choose a sampling grid that is a multiple of the Nyquist rate, select the columns of the Fourier matrix that correspond to the $b$ coefficients needed in Lemma \ref{lem:dft_cs_for_wD_series}, giving a tall matrix, and then consider the measurement matrix that is a random sub-selection of rows from this tall matrix. Going through such an analysis results in an identical scaling in the number of measurements as stated in~\Cref{thm:subsampled_3ddft} and the remarks following it. Thus, we can investigate the performance of the 2DDFT CS method as the grid density increases.

\begin{figure}[ht]
    \centering
    \includegraphics[width=\reprintcolumnwidth]{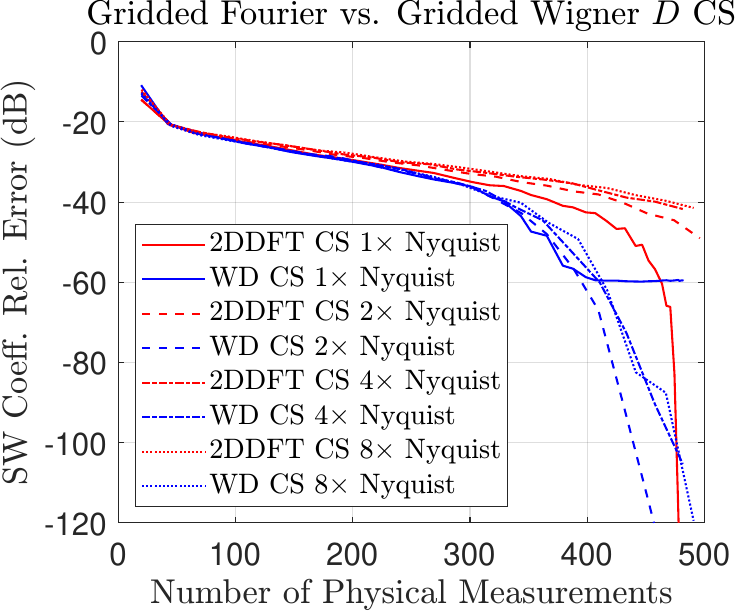}
    \caption{\tit{Coefficient Relative Error for On-Grid Fourier CS versus On-Grid Wigner $D$-function CS.} Averaging over 25 trials for each number of measurements, the proposed Fourier-based CS performs slightly worse than the on-grid Wigner $D$-function CS when sampling from the Nyquist grid. When the grid density increases, the performance of both methods degrade slightly. Note, the on-grid Wigner $D$-function approach does not have any theoretical backing like the Fourier method does in~\Cref{thm:subsampled_3ddft}.}
    \label{fig:on_grid_D_vs_F}
\end{figure}

To that end, we compare the 2DDFT CS method as a function of the number of unique ``physical measurements'' and increasing sample grids. It is also interesting to include in this comparison a similarly sampled Wigner $ D$-function-based approach. In particular, although there is no {\em a priori} reason to believe using a Wigner $D$-function measurement matrix sampled on the Nyquist grid should work, we can test to see if using the same samples as the proposed CS method gives reasonable results. To compare these we first reconstruct \tbf{C1a} by first randomly selecting measurements from the Nyquist grid and integer multiples of it on the sphere (i.e., $n$ times as many available samples spatially in each direction). We use these samples to carry out the 2DDFT CS method. Then we reconstruct the \tbf{C1a} SW coefficients by randomly selecting the same number of ``physical measurements'' from the same sample grids and carry out BP using a Wigner $D$-function measurement matrix with $\mu = 0$ functions only (i.e., rescaled spherical harmonics) and the appropriate preconditioning from~\cite{bangunSensingMatrixDesign2020}. In the continuous case, the preconditioning is so that samples can be uniformly selected from the domain, so we heuristically view this approach as discrete samples from the appropriate uniform continuous case. The results of this can be seen over the range of possible measurements in~\Cref{fig:on_grid_D_vs_F}, where we average over 25 trials at each number of measurements. For small sample numbers from all grid densities, the 2DDFT CS approach performs comparably to the gridded Wigner $D$-function approach. Near full sampling, the gridded Wigner $D$-function approach performs slightly better. The reasoning for this is the same as seen in the continuous case presented in the text related to~\Cref{fig:recon_error_versus_meas}. Interestingly, when the measurement number is closer to full sampling, increasing the grid density causes an increase in relative error for both methods. We discuss this worsening of the 2DDFT CS method in the next section. As a note, we would like to emphasize that the on-grid Wigner $D$-function approach does not provide a theoretical guarantee like \Cref{thm:subsampled_3ddft} does for the 2DDFT CS method. Thus, its usefulness cannot be guaranteed to extend to other coefficients nor can we guarantee that it will degrade gracefully with noise.

\subsection{SW Coefficient Recovery in the Presence of Noise}
\begin{figure}[ht]
        \centering
        \includegraphics[width =\reprintcolumnwidth]{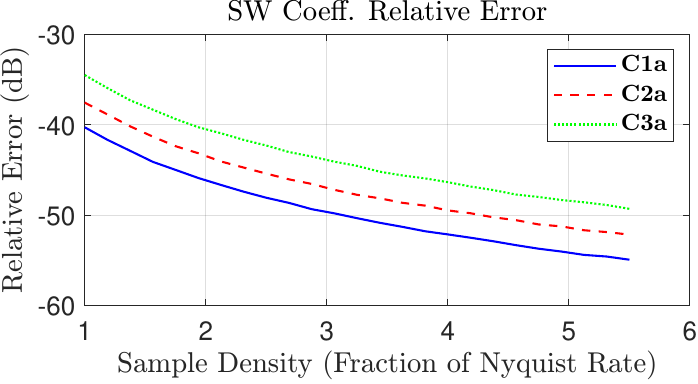}
        \caption{\tit{Classical Fourier Reconstruction with Measurement Noise.} SW coefficients from \tbf{C1a}, \tbf{C2a}, and \tbf{C3a} reconstructed according to~\cite{wittmannProbecorrectedSphericalNearfield1992} with zero-mean Gaussian measurement noise whose variance is $40$ dB below the maximum field value in each case. With a band-limit of $n_{\max}=15$, the Nyquist rate sampling requires 496 measurements.}\label{fig:fourier_w_noise}
\end{figure}

The examples above all contained no measurement noise. In the presence of measurement noise, even classical Nyquist sampling approaches will have their performance degrade. However, in order to improve accuracy in the presence of noise, oversampling at a rate greater than the Nyquist rate is common. For example, we simulate measurements of the sound fields generated by \tbf{C1a}, \tbf{C2a}, and \tbf{C3a} with mean zero and variance that is $40$ dB below the peak value for each case. Using these simulated measurements then we plot the coefficient relative error, \eqref{eq:coef_rel_err}, resulting from the classical fully sampled Fourier method in~\cite{wittmannProbecorrectedSphericalNearfield1992}. The results are given in \Cref{fig:fourier_w_noise}. As can be seen in \Cref{fig:fourier_w_noise}, increasing the sampling to five times the Nyquist rate results in a decrease in the relative error of nearly 15 dB for each case. Note the curves are split apart because the peak field value, and thus total noise, increases from \tbf{C1a} to \tbf{C2a} and again to \tbf{C3a}. 
    
\begin{figure}[ht]
    \centering
    \figcolumn{
        \fig{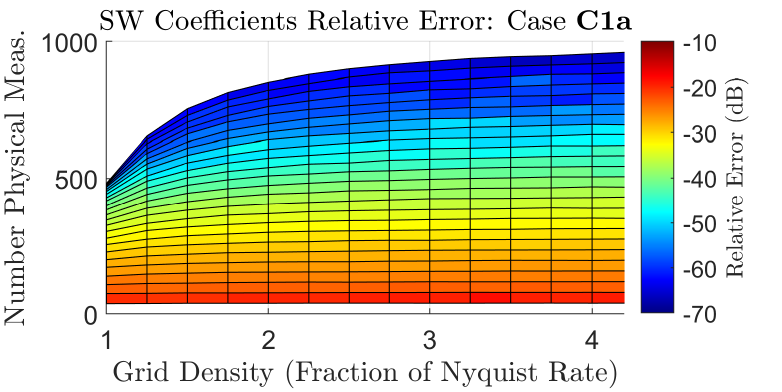}{\reprintcolumnwidth}{(a)}
        \fig{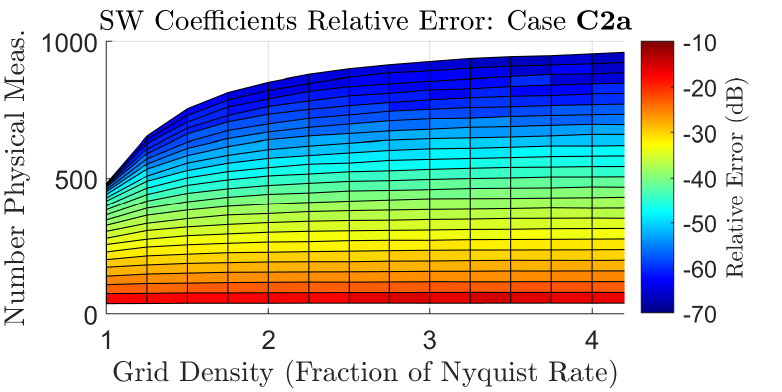}{\reprintcolumnwidth}{(b)}
        \fig{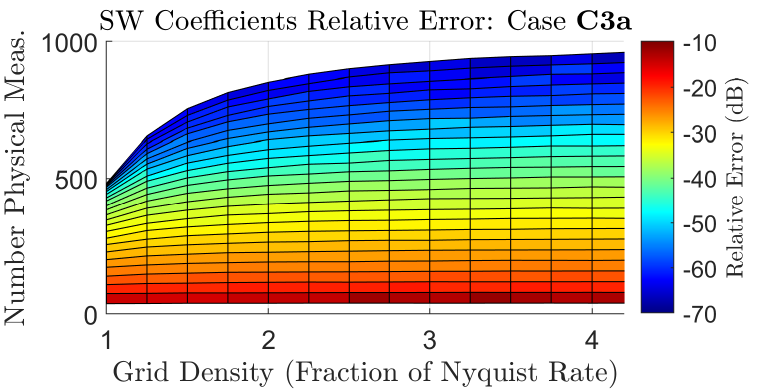}{\reprintcolumnwidth}{(c)}
    }
    \caption{{\it CS Coefficient Relative Error Versus Number of ``Physical Measurements'' in Cases \tbf{C1a, C2a, and C3a.}} The 2DDFT CS method for cases \tbf{C1a}, \tbf{C2a}, and \tbf{C3a} with added zero-mean Gaussian measurement noise with variance $40$ dB below the maximum field value. For each case, 2DDFT CS with a fixed number of samples and increased grid density results in slightly degraded relative error. Note, at the band-limit of $n_{\max}=15$, the Nyquist grid has 496 possible measurements.}\label{fig:cs_recovery_vs_grid_density}
\end{figure}

\begin{figure}[ht]
    \centering
    \includegraphics[width =\reprintcolumnwidth]{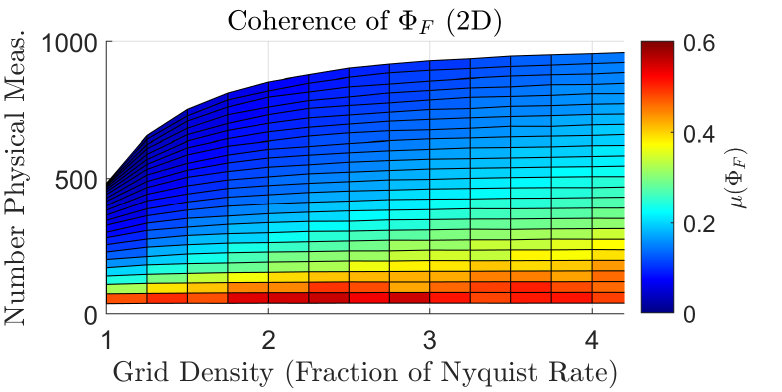}
    \caption{{\it Coherence of $\Phi_F$ in 2D.} The average coherence of the 2DDFT CS measurement matrix is suggestive of decreased performance for increased sample grid density and fixed measurement number. Averaging is taken over 25 trials for each combination of grid density and measurement number.}\label{fig:coherence_vs_grid_density}
\end{figure}

As discussed in the previous section, we can increase the sample grid in the 2DDFT CS method and use it under the same guarantees as seen in~\Cref{thm:subsampled_3ddft}. This allows us to directly compare our method with those results in~\Cref{fig:fourier_w_noise}. To that end, we first investigate the relative error from CS reconstructions of the SW coefficients in \tbf{C1a}, \tbf{C2a}, and \tbf{C3a} with the $-40$ dB Gaussian noise used in~\Cref{fig:fourier_w_noise} as a function of the number of ``physical measurements'' and the grid density. For each sampling number and grid density, we average the relative error over 25 trials in~\Cref{fig:cs_recovery_vs_grid_density}. As can be seen in~\Cref{fig:cs_recovery_vs_grid_density}, for a fixed number of measurements, the relative error slightly degrades as the grid density increases. One might assume the accuracy should be constant with the fixed number of measurements. However, this is not the case. In particular, for a fixed number of ``physical measurements'', the likelihood of the measurement matrix being worse for CS increases as the sample grid increases. This is visualized in \Cref{fig:coherence_vs_grid_density}, where we plot the coherence of the column normalized version of $\Phi_F$ for the 2DDFT CS method. The coherence of a matrix $\Phi \in \mbb{C}^{M\times N}$ with normalized columns is defined as~\cite{foucartMathematicalIntroductionCompressive2013}
\begin{equation}
    \mu(\Phi) = \max_{1 \leq i\neq j \leq N} \left| \phi_i^H \phi_j \right|,
\end{equation}
where $\phi_i$ is a column of $\Phi$.
To get physical intuition for this decrease in performance we hypothesize the following. If we sample near the Nyquist rate then the recovery should be near perfect, but if the grid density increases, then measurements have an increased likelihood to cluster in certain locations becoming more coherent and leaving larger gaps of the field unmeasured. We note this is a hypothesis since clustering measurements is a statement about the rows of the measurement matrix, while the coherence is a measure of column-wise relationships, and so rigorous statements about the relationship are likely to be much more subtle.

\begin{figure}[ht]
    \centering
    \figcolumn{
        \fig{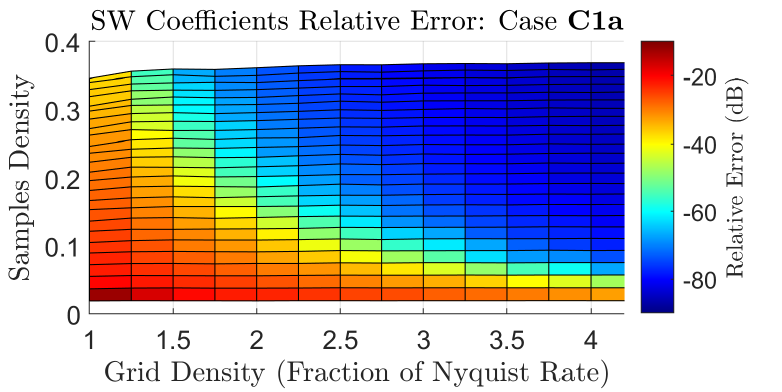}{\reprintcolumnwidth}{(a)}
        \fig{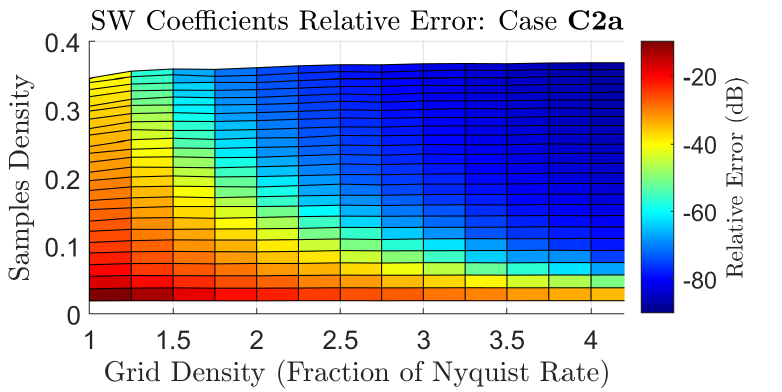}{\reprintcolumnwidth}{(b)}
        \fig{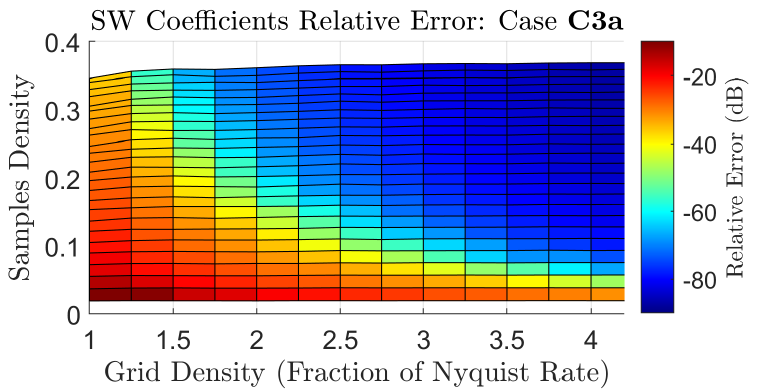}{\reprintcolumnwidth}{(c)}
    }
    \caption{{\it CS Coefficient Relative Error Versus Sample Grid Density in Cases \tbf{C1a, C2a, and C3a.}} with additive measurement noise, fixing the sample density (measurment \# divided by \# of possible measurements) and increasing grid density for cases \tbf{C1a}, \tbf{C2a}, and \tbf{C3a} results in improved relative error using the 2DDFT CS method. In other words, increasing the sampling rate and sample number improves the denoising achieved by the 2DDFT CS method and outperforms oversampling with the classical method in~\Cref{fig:fourier_w_noise}. The Noise used here is added zero-mean Gaussian measurement noise with variance $40$ dB below the maximum field value. The band-limit used, $n_{\max}=15$, results in the Nyquist grid having 496 possible measurements.    }\label{fig:cs_recovery_vs_sample_density_vs_grid_density}
\end{figure}

The intent of increasing the sampling rate in the classical Fourier approach to recover the SW coefficients in~\cite{wittmannProbecorrectedSphericalNearfield1992} is to increase the method's robustness to noise. \Cref{fig:cs_recovery_vs_grid_density} showed that increasing grid density and keeping the sample number the same does not improve performance when using CS with Lemma \ref{lem:dft_cs_for_wD_series}. However, if we consider fixed sampling densities (number of measurements used divided by the total possible number) as the density of the sampling grid is increased, we see improvements in our proposed CS approach. \Cref{fig:cs_recovery_vs_sample_density_vs_grid_density} shows these results for \tbf{C1a} to \tbf{C2a} and again to \tbf{C3a} with the same noise as before. Similar to the classical Fourier case, for a fixed sample density and increased grid density, the relative error decreases for each case tested. This can be interpreted as de-noising that occurs by promoting more sparse coefficients in QCBP. Interestingly, the de-noising from the proposed CS approach with a denser than Nyquist grid and sub-sampling gives better results than the de-noising attained from using oversampling with the method from~\cite{wittmannProbecorrectedSphericalNearfield1992}. This indicates, at least in this case, the de-noising benefits of CS via QCBP are an added benefit beyond simply decreasing the required number of measurements. Thus, if one is currently using Nyquist sampling at some denser grid than Nyquist, using the CS approach \emph{and fewer measurements} may improve accuracy. For example, CS with a sample density of nearly 1/3 at two times the Nyquist rate beats the classical Fourier approach with sample density 1  at two times the Nyquist rate by nearly 20 dB or more in \tbf{C1a}, \tbf{C2a}, and \tbf{C3a}.

\section{Conclusion} \label{sec:conclusion}
We have developed an approach to recover SW or SH expansion coefficients using compressive samples taken from a pre-defined grid. This approach not only avoids using measurements at arbitrary positions on the sphere or $\sot$, as is common for BOSs, but it does so while maintaining robust reconstruction guarantees. For sufficiently sparse signals in both domains, the number of measurements required for robust reconstruction have sub-linearly scalings with the basis dimension. These scalings are slightly worse than the best cases from the literature~\cite{burqWeightedEigenfunctionEstimates2012, bangunSensingMatrixDesign2020}, however, these references methods require samples from arbitrary positions on the sphere and $\sot$.

Using our proposed CS approach, we numerically compared its results with a commonly used Fourier approach to recover SW or SH expansion coefficients. In our tests, our on-grid CS approach performed comparably in the presence of noise when a Nyquist grid is used. However, when the grid sampling is increased to two times the Nyquist rate, our CS approach boasted increased de-noising capabilities while using fewer measurements. For the three examples used to test the CS approach, the relative error for the CS method beats the classical Fourier approach by 20 dB or better. Moreover, this was achieved while using a third of the measurements needed for the classical Fourier approach.

The CS approach developed in this paper allows for field reconstructions in various application areas like acoustic spherical holography, loudspeaker characterizations, and even EM antenna characterizations. In some of these cases, measurements are restricted to certain areas on the sphere or $\sot$~\cite{valdezRestrictedDomainCompressive2022}. The work in ~\cite{valdezRestrictedDomainCompressive2022} can straightforwardly be generalized to the approach developed in this paper so that gridded and restricted measurements can be used for SW or SH field reconstructions. However, the effect of compounding transformations (continuous Wigner $D$-function to discrete Fourier and then to a Slepian basis) on the conservation of sparsity throughout the bases requires further investigation and would be a suitable future direction of study.





\appendix*
\section{Proof of Lemma 6} \label{app:ProofOfMainLem}
First, we note that the Wigner $d$-function with integer $n$ is periodic on $2\pi$~\cite{varshalovichQuantumTheoryAngular1988} and can be expressed as a Fourier series. This Fourier series is band-limited and has the form~\cite{hansenSphericalNearfieldAntenna1988}
\begin{equation}\label{eq:wigner_d_fourier}
    \t{d}{_n^{\mu m}}(\beta) = \mi^{\mu-m}\sum_{m'=-n}^n \t{\Delta}{_n^{m', \mu}}\t{\Delta}{_n^{m',m}} e^{-\mi m' \beta},
\end{equation}
where
\begin{align}
    \t{\Delta}{_n^{m',\mu}} &= \t{d}{_n^{m' \mu}}\left(\frac{\pi}{2}\right),\\
    \t{\Delta}{_n^{m',m}} &= \t{d}{_n^{m' m}}\left(\frac{\pi}{2}\right).
\end{align}
Substituting \eqref{eq:wigner_D} and \eqref{eq:wigner_d_fourier} into \eqref{eq:wigner_D_expanison} gives
\begin{align}
    w_j & = \sum_{n=0}^{n_\text{max}}\sum_{m,\mu,m'=-n}^n \t{\chi}{_n^{m \mu}} e^{-\mi m \gamma_j -\mi \mu \alpha_j - \mi m' \beta_j} + \eta_j, \\
    \t{\chi}{_n^{m \mu}} & = \mi^{\mu-m} \t{\Delta}{_n^{m',\mu}}\t{\Delta}{_n^{m',m}}  \acoef{n}{m\mu}.
\end{align}
We now reorder the sums so that the sum over $n$ is on the inside, yielding
\begin{equation}
    w_j = \sum_{m,\mu,m'=-n_{\max}}^{n_{\max}}  \sum_{n=n_{\min}}^{n_{\max}} \t{\chi}{_n^{m \mu}}  e^{-\mi m \gamma_j -\mi \mu \alpha_j - \mi m' \beta_j} + \eta_j,
\end{equation}
where $n_{\min}=\max(|m|,|\mu|,|m'|)$. In anticipation of benefits in terms of measurements (see ~\Cref{rem:fourier_grid_size}), we extend the ranges of the $m', m,$ and $\mu$ sums to range from $n_{\max-1}$ to $n_{\max}$, giving
\begin{equation}\label{eq:reordered_extended_index}
    w_j = \sum_{m,\mu,m'=-n_{\max}-1}^{n_{\max}}  \sum_{n=n_{\min}}^{n_{\max}} \t{\chi}{_n^{m \mu}}  e^{-\mi m \gamma_j -\mi \mu \alpha_j - \mi m' \beta_j} + \eta_j,
\end{equation}
where we define $\t{\Delta}{_n^{m', m}}$ and $\t{\Delta}{_n^{m', \mu}}$ to be zero if $m'$, $m$, or $\mu$ is $-n_{\max}-1$.
Next, define the sum over $n$ in \eqref{eq:reordered_extended_index} as 
\begin{equation}
    \bcoef{m'}{m\mu} = \sum_{n=n_{\min}}^{n_{\max}} \t{\chi}{_n^{m \mu}}.
\end{equation} 
Thus, we arrive at a restatement of our problem in two parts. First, we solve for the coefficients $\bcoef{m'}{m\mu}$ from the linear problem in \eqref{eq:fourier_expansion},
\begin{equation}
    w_j = \sum_{m,\mu,m'=-n_{\max}-1}^{n_{\max}}  \bcoef{m'}{m\mu} e^{-\mi m \gamma_j -\mi \mu \alpha_j - \mi m' \beta_j}  + \eta_j. \label{eq:fourier_expansion}
\end{equation}
This amounts to finding the Fourier series coefficients of $w$, $\bcoef{m'}{m\mu}$, from a set of measurements. Then, we solve for the Wigner $D$-function coefficients $\acoef{n}{m\mu}$ from the Fourier coefficients $\bcoef{m'}{m\mu}$ using the following linear inverse problem:
\begin{equation}
    \begin{split}
        \forall  m, \mu &\in \{-n_{\max}-1,-n_{\max},\cdots,n_{\max}\}\\
        \bcoef{m'}{m\mu} &= \sum_{n=n_{\min}}^{n_{\max}} \t{\chi}{_n^{m \mu}}, \\
        \t{\chi}{_n^{m \mu}} &= \mi^{\mu-m} \t{\Delta}{_n^{m',\mu}}\t{\Delta}{_n^{m',m}}  \acoef{n}{m\mu}.
    \end{split}
     \label{eq:auxiliary_problem}
\end{equation}
These equations can be equivalently written as matrix equations
\begin{align}
    w &= \Phi_F b + \eta, \label{eq:fourier_problem}\\
    \begin{split}
        \forall m, \mu &\in \{-n_{\max}-1,-n_{\max},\cdots,n_{\max}\}, \\
        \bcoef{}{m \mu} &= \t{B}{^{m\mu}} \acoef{}{m \mu} \label{eq:a_b_coef_problem},
    \end{split}
\end{align}
where $\Phi_F$ is $M\times N_F$ with $N_F$ being the number of band-limited complex exponential functions $N_F = (2n_{\max}+2)^3$. Here we have used
\begin{align}
    \left[\Phi_F\right]_{i,j} &= e^{-\mi m(j) \gamma_i} e^{-\mi \mu(j) \alpha_i} e^{-\mi m'(j) \beta_i} \\
    \til{n}_{\min} &= \max(|m|,|\mu|), \\
    \acoef{}{m \mu} &= [\acoef{\til{n}_{\min}}{m\mu}, \acoef{\til{n}_{\min}+1}{m\mu},\cdots,\acoef{n_{\max}}{m\mu}]^T, \\
    \bcoef{}{m \mu} &= [\bcoef{-n_{\max}}{m\mu}, \bcoef{-n_{\max}+1}{m\mu}, \cdots, \bcoef{n_{\max}}{m\mu}]^T,
\end{align}
for some ordering $m(j)$, $\mu(j)$, $m'(j)$, with the vector $b$ correspondingly arranged. 
The matrices $\t{B}{^{m\mu}} \in \mbb{C}^{\dim_1\times \dim_2}$ with $\dim_1=2n_{\max}+2$ and $\dim_2=n_{\max}+1-\til{n}_{\min}$ have elements
\begin{equation}\label{eq:B_m_mu_entires}
    \begin{split}
        \left[\t{B}{^{m\mu}}\right]_{i,j} & = \begin{cases}
            \t{\psi}{^{m \mu}},& |i-n_{\max}-1| \leq \til{n}_{\min}-1+j \\
            0,&  \text{otherwise}
        \end{cases},\\
        \t{\psi}{^{m \mu}} &= \mi^{\mu-m}\t{\Delta}{_{\til{n}_{\min}-1+j}^{i-n_{\max}-1, \mu}}\t{\Delta}{_{\til{n}_{\min}-1+j}^{i-n_{\max}-1, m}}
    \end{split}
\end{equation}
where $i=1,2,\cdots,2n_{\max}+2$ and $j=1,2,\cdots,n_{\max}+1-\til{n}_{\min}$. The size of the matrix $\t{B}{^{m\mu}}$ can be seen from the fact that there are no $\acoef{n}{m \mu}$ with $n<\til{n}_{\min}$. The restriction on the $\t{B}{^{m\mu}}$ values can be seen from the fact that $\t{\Delta}{_n^{m',\mu}}\t{\Delta}{_n^{m',m}}=0$ when $|m'|\geq n$.

Now select possible measurement points given by $(\alpha_j, \beta_k, \gamma_l) = (2\pi j/(2n_{\max}+2), 2\pi k/(2n_{\max}+2), 2\pi l/(2n_{\max}+2))$ for $j,k,l\in\{-n_{\max}-1,-n_{\max},\cdots,n_{\max}\}$. We call this the Nyquist grid on $\tthre$. Thanks to the double covering of $\sot$ by $\tthre$, the Nyquist grid on $\tthre$ straightforwardly maps to a grid of measurement points on $\sot$. Rewriting \eqref{eq:fourier_expansion} with these selected measurement points gives 
\begin{equation} \label{eq:3ddft_series}
    \begin{split}
        w_{jkl} & = w(\alpha_j,\beta_k,\gamma_l) \\
        & = \sum_{\mu,m,m'=-n_{\max}-1}^{n_{\max}}  \bcoef{m'}{m\mu} e^\frac{-\mi 2\pi (\mu j+ m k+m' l)}{2n_{\max}+2} + \eta_{jkl}.
    \end{split}
\end{equation}
This can be recognized as the 3DDFT of the coefficients $\bcoef{m'}{m\mu}$. Thus, if we sample at a subset $\Omega$ of all of these possible positions, we have the matrix problem,
\begin{align}
    N_F^{-1/2} w &= P_\Omega U_F b + N_F^{-1/2} \eta,
\end{align}
where $N_F = (2n_{\max}+2)^3$, $U_F \in \mbb{C}^{N_F \times N_F}$ is the unitary matrix representing the 3DDFT, and $P_\Omega$ is the matrix selecting the subset of rows $\Omega$ of $U_F$.
\bsquare




\bibliography{references.bib}




\end{document}